\documentclass[aps,amsmath,amssymb,reprint]{revtex4-1}
\usepackage{graphicx}
\usepackage{amsmath}
\usepackage{dcolumn}
\usepackage{soul}
\usepackage{cancel}
\usepackage{bm}
\usepackage[nice]{nicefrac}
\usepackage[utf8]{inputenc}
\usepackage[T1]{fontenc}
\usepackage{mathptmx}
\usepackage{dsfont}
\usepackage[dvipsnames]{xcolor}
\usepackage[normalem]{ulem} 
\usepackage{etoolbox}
\usepackage{amssymb}
\DeclareUnicodeCharacter{0301}{\'{e}}

\onecolumngrid

\begin{document}
\preprint{APS/123-QED}
\title[MC objects in 1D]{Diffusive spreading of a polydisperse polymer solution in a channel}
\author{H. Y. Wang}
\email{hwang188@uottawa.ca}

\author{G. W. Slater}
\email{gary.slater@uottawa.ca}
\affiliation{Department of Physics, University of Ottawa,~Ottawa, Ontario K1N 6N5, Canada}

\date{\today}

\begin{abstract}
Long DNA molecules can be mapped by cutting them with restriction enzymes inside a narrow channel. Once cut, the individual fragments thus produced move away from each other due to diffusion and entropic effects. We investigate how long it takes for these fragments to travel distances large enough for an experimental device to distinguish them and (possibly) estimate their size. In essence, this is a single-file diffusion process in which molecules of different sizes and hence different diffusion coefficients spread out from an initially dense configuration. We use Monte Carlo methods to investigate this class of problems and define the time taken to reach the required final state as a first-passage \textit{spreading time}. Our results demonstrate that the stochastic nature of the diffusion process is as significant as the specifics of the molecular size distribution in determining the spreading time. We examine the relationship between the spreading time and the final space occupied by the fragments as a function of the experimental parameters and determine the fundamental length scale governing this process. We introduce a molecular sequence randomness parameter, $Z$, which is linearly correlated with the final spreading time. Finally, we show that the distribution function of spreading times follows a well-known form for first-passage time problems, and that its variance decreases linearly with the number of fragments. 


\end{abstract}

\maketitle

\section{\label{sec:intro}Introduction}

Nanofluidic channels are becoming very powerful instruments in laboratories~\cite{ref:dorfman_review}, especially for studying and analyzing DNA molecules. For example, Schwartz et al.~\cite{ref:Schwartz1,ref:Schwartz2} introduced the idea of using nanochannels to map long DNA molecules~\cite{ref:melt,ref:denseandsparse}. Since then, the methodologies used for DNA imaging and data analysis, as well as the implementation of mapping techniques, have advanced \cite{ref:Ebenst_review,ref:Wang_review}.

This approach to DNA mapping typically involves extending a long DNA chain within a narrow channel, cutting it using restriction enzymes \cite{ref:Mapping_review,ref:Schwartz1}, and imaging the distribution and order of the resulting fragments in the channel~\cite{ref:channelCUtting}. Restriction enzymes target short, specific DNA sequences and cut the DNA chain into smaller fragments in a very precise and reproducible way\cite{ref:enzyme_cut1,ref:enzyme_cut2}. Because the cutting process is performed within a narrow channel, the order of the restriction fragments is maintained (fragments cannot change position due to excluded volume interactions), allowing molecular mapping once the fragments can be imaged. 

The objective of this study is to explore a particular dimension of this problem. Specifically, we are interested in investigating the dynamics of a linear polymer chain confined within a channel after it has been cut into several fragments. The fundamental question we seek to address is the time required for these fragments to become visibly separated from each other to a degree that allows for their unambiguous detection. This process is influenced by two key factors: entropic repulsion and diffusion. We will focus on understanding the parameters that control the necessary separation time; in doing so, we will use parameters that roughly match the conditions used for DNA mapping in channels (e.g. $N=5-20$ fragments) while also investigating the (admittedly more theoretical) asymptotic limits (e.g., $N \gg 1$). Of course, the size and order of the molecular fragments are also important factors; for example, very large fragments can block the diffusion process of smaller ones if located near the ends of a sequence of fragments. 

The paper is organized as follows.  In Section \ref{sec:Theory}, we describe our Monte Carlo model and the necessary parameters for analyzing simulation data. Section \ref{sec:Results} presents our results in several subsections covering the following topics: the properties of monodisperse systems; a study of some simplified bimodal systems; a comparison between random and monodisperse systems, leading to the introduction of a parameter $Z$ to characterize the molecular arrangement; the role of permutations in random systems; and finally, a look at extreme situations and distribution functions of spreading times. Our conclusions, including possible future extensions of these results, are found in Section \ref{sec:conclution}.

\section{\label{sec:Theory}The Lattice Monte Carlo Model}

\begin{figure}[!ht]
\centering
\includegraphics[width = 0.482\textwidth]{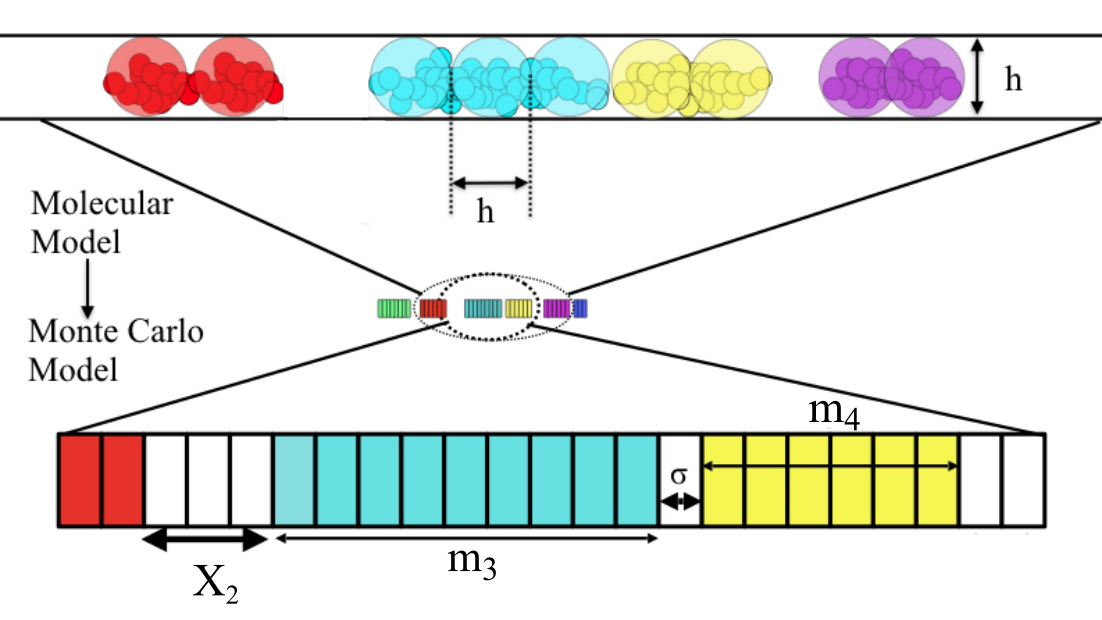}
\caption{Diagram of the molecular model and its mapping onto the simulated system. Top panel: A long polymer chain inside a narrow tube of diameter $h$ was cut into 4 fragments. The fragments form blobs of size $h$ aligned along the tube axis. Middle panel: The system is then mapped onto a one-dimensional Monte Carlo model with rigid objects. Bottom panel: The $i^{th}$ object comprises $m_i \ge 1$ lattice cells of size $\sigma=1$. The target state is reached when all gaps $X_i$ are simultaneously larger than a threshold value $\tilde{X}$ for the first time.}
\label{fig:system}
\end{figure}

The top part of Fig.~\ref{fig:system} shows a molecular view of the system. A long polymer chain of mass $M$ is cut into $N$ pieces at time $t=0$, after which the pieces start to move away from each other. The diameter of the channel, $h$, is smaller than the free-solution radius of gyration $R_g$ of all the pieces; as a result, the molecular fragments form blobs of diameter $h$ that are aligned along the channel axis. We denote by $\tilde{X}$ the minimum gap size that can be detected between two adjacent fragments; in other words, the distance between two fragments must be $\ge \tilde{X}$ so that we can confirm the presence of two fragments instead of a single, longer one. For example, this minimum gap may have to be of order $h$ (e.g., the $X_2$ gap in Fig.~\ref{fig:system}).

We map this molecular system onto a one-dimensional Lattice Monte Carlo model, also shown in Fig.~\ref{fig:system}. The lattice cell size, $\sigma=1$, is smaller than both $h$ and the smallest gap, $\tilde{X}$. The system thus consists of $N$ rigid molecules diffusing in the single-file diffusion (SFD) regime; the fluctuations in the axial length of the molecules are neglected. Because hydrodynamic interactions are screened between blobs \cite{ref:deGennes_1,ref:deGennes_2}, we use a Rouse model for the polymer chains; the diffusion coefficient of molecule $i$ is then $D(m_i)=D_1/m_i$, where $m_i \ge h$ refers to its size (\textit{i.e.}, number of lattice cells occupied). The total molecular size is thus $M=\sum_{i=1}^N m_i$. We use the optimized Monte Carlo moves described in ref.~\cite{ref:MykytaV2012}.

We denote by $X_i~(i=1..N-1)$ the spacing or gap between molecules $i$ and $i+1$. At time $t=0$, the molecules are equally spaced by a single lattice cell: $X_i(0)=1,~\forall~i$. We simulate the evolution of this system until \textit{all} gaps are simultaneously greater than a certain target value $\tilde{X}$ \textit{for the first time}, 
\begin{equation}
   X_i(t=\tau_\mathsf{\,f}) \ge \tilde{X}~~~~~~~~~\forall~i ~,
\end{equation}
which defines the \textit{spreading time} $\tau_\mathsf{\,f}$. Our main goal is to understand how the mean first passage time $\tau_\mathsf{\,f}$ depends on the total molecular size $M$, the number of molecular fragments, $N$, the minimum final gap size, $\tilde{X}$, and especially the details of the distribution of molecular sizes, $(m_i) \equiv (m_1,m_2,..,m_N)$.

In addition to the spreading time $\tau_\mathsf{\,f}$, the span of the system (i.e., the difference between the largest and smallest axial coordinates of the $M$ lattice cells that constitute the molecular fragments), $S(t)=M+\sum_{i=1}^{N-1}~X_i(t)$, is also of interest. As we chose $X_i(0)=1$, its initial value is $S(0)=M+N-1$. We will look at the mean increase of the span, $\Delta S_\mathsf{\,f}  = \left< S(\tau_\mathsf{\,f})-S(0) \right> = \sum_{i=1}^{N-1}~[X_i(\tau_\mathsf{\,f})-1]$, during the process. The minimum value of $\Delta S_\mathsf{\,f}$ compatible with the desired final state is
\begin{equation}
   \Delta \, \mathbf{s} \equiv \min \left[ \Delta S_\mathsf{\,f}\right]=(N-1)(\tilde{X}-1).
\end{equation}
As we shall see, $\Delta \, \mathbf{s}$ turns out to be a key length scale. Finally, the links between $\Delta S_\mathsf{\,f}$, $\tau_\mathsf{\,f}$ and the final mean square center-of-mass (\textsc{cm}) displacement of the population, $\left< \Delta x_{\textsc{cm}}^2(\tau_\mathsf{\,f}) \right>$, are studied to understand what controls $\tau_\mathsf{\,f}(N,h,(m_i))$.

\textcolor{black}{As usual with lattice Monte Carlo algorithms, we displace only one randomly-selected particle at a time. However, when the particles have different sizes $m_i$, the rate at which they diffuse is not uniform since $D(m_i)=D_1/m_i$. The probability of choosing a given molecule is then $W_i \sim 1/m_i$; when normalized, the probability reads $W_i=m_i^{-1}/\sum_{i}{m_i^{-1}}$.}

The (dimensionless) simulation data will be given in units of $\sigma$ for distances and $\tau_1 = \sigma^2/2D_1$, where $D_1 \equiv D(m=1)$, for times. The ensemble averaged results will be denoted $A{(a)}$, where $a$ is the statistical error on the mean. In some cases, we will use the notation $A[\alpha]$ to give the standard-deviation $\alpha$ of the distribution of values for $A$.

\section{\label{sec:Results}Results}

\subsection{Monodisperse solution of molecules}
\label{sec:mono}

\begin{figure}[!ht]
\centering
\includegraphics[width = 0.482\textwidth]{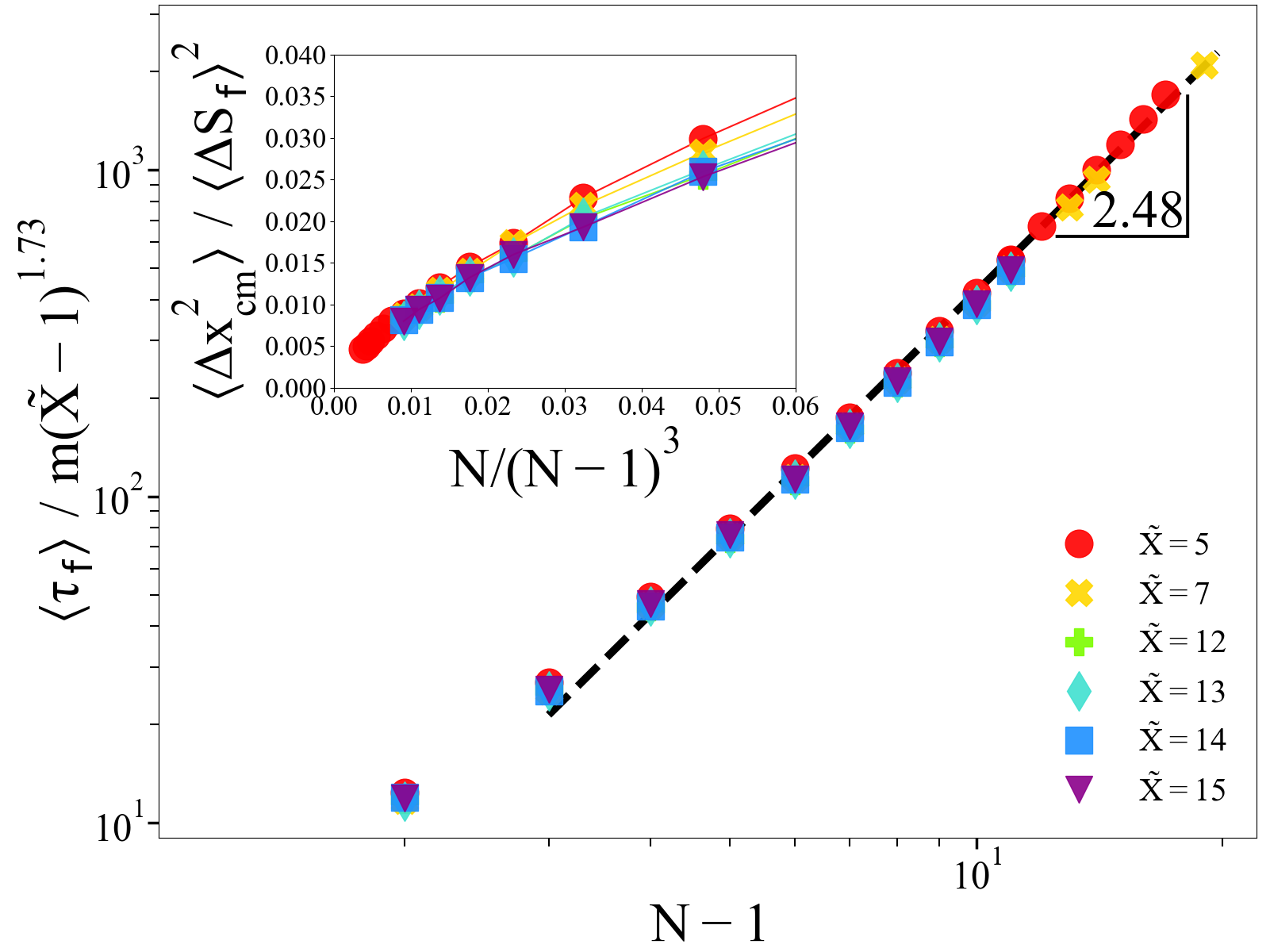}
\caption{Scaled mean spreading time $\langle \tau_\mathsf{\,f} \rangle/m(\tilde{X}-1)^{1.73}$ \textit{vs}. $N-1$ for an ensemble of $10\,000$ monodisperse systems made of $N$ molecules of size $m$. The data for different final gaps $\tilde{X}$ collapse; the linear fit uses the range $N \in [10,20]$. The exponent 1.73 was found to collapse the data for the different values of $\tilde{X}$. Inset:  Relative displacement ratio $\langle \Delta x_{\textsc{cm}}^2(t=\tau_\mathsf{\,f}) \rangle /\langle \Delta S_\mathsf{\,f}\rangle^2$ \textit{vs.} $N/(N-1)^{3}$. }
\label{fig:powerlaws}
\end{figure}

We first study how a group of identical molecules of size $m$ reaches the desired final state. Since $D \sim 1/m$, the value of $m$ only affects the spreading time via a trivial pre-factor. We thus expect a scaling law of the form
\begin{equation}
\label{eq:tauf_scaling}
    \langle \tau_\mathsf{\,f}(N,\tilde{X}) \rangle \sim m~(N-1)^{\eta_\tau}~(\tilde{X}-1)^{\chi_\tau}~,
\end{equation}
where the $(N-1)$ comes from the fact that we have $N-1$ gaps growing simultaneously, while the $(\tilde{X}-1)$ represents the necessary increase in gap size. Our results (see Fig.~\ref{fig:powerlaws}) for the ranges $N \in [10,20]$ and $\tilde{X} \in [8,20]$ are consistent with
\begin{equation}
    \langle \tau_\mathsf{\,f} \rangle \sim m~(N-1)^{2.48(2)}~ (\tilde{X}-1)^{1.73(1)}~.
\label{Eq:tau_f_value}
\end{equation}
Extrapolating our results (data not shown), we find that ${\eta_\tau}(N\to \infty) \approx 2.93(5)$ and ${\chi_\tau}(\tilde{X} \to \infty) \approx 1.99(4)$, suggesting the asymptotic behavior
\begin{equation}
   \langle \tau_\mathsf{\,f} \rangle \sim m~N^3 \tilde{X}^2 \sim M \times \left[\Delta \, \mathbf{s}\right]^2 \sim \left[\Delta \, \mathbf{s}\right]^2/ \, D(M).
\label{eq:taufdM}
\end{equation}
Interestingly, the mean spreading time scales like the time required for the uncut molecule of size $M$ to diffuse over a distance equal to the minimum span increase compatible with the desired final state, $\Delta \, \mathbf{s}$. We also note that $\langle \tau_\mathsf{\,f} \rangle \sim M\,N^2 \sim m\, N^3$ increases very rapidly with the number of fragments.

Similarly, the final mean square displacement of the \textsc{cm} was fitted using the scaling form
\begin{equation}
    \langle \Delta x_{\textsc{cm}}^2(\tau_\mathsf{\,f}) \rangle \sim N(N-1)^{\eta_x}~ \,(\tilde{X}-1)^{\chi_x}~.
\end{equation}
Our results give (data not shown)
\begin{equation}
    \langle \Delta x_{\textsc{cm}}^2(\tau_\mathsf{\,f}) \rangle \sim N  \,(N-1)^{0.6(1)}~(\tilde{X}-1)^{1.8(1)}~,
\label{Eq: size_dep}
\end{equation}
as well as ${\chi_x}(\tilde{X} \to \infty)  \approx {2.0(1)}$ and ${\eta_x}(N\to \infty) \approx 0.9(1)$ for an asymptotic limit 
\begin{equation}
  \langle \Delta x_{\textsc{cm}}^2(\tau_\mathsf{\,f}) \rangle \sim N^2 \, \tilde{X}^{2} \sim \left[\Delta \, \mathbf{s}\right]^2. 
\label{Eq:dx2_period}
\end{equation}
The $\textsc{cm}$ of the group diffuses over a distance comparable to the minimum increase $\Delta \, \mathbf{s}$, consistent with Eq.~\ref{eq:taufdM}.

Finally, the increase of the span was fitted using the form
\begin{equation}
    \Delta S_\mathsf{\,f} \sim (N-1)^{\eta_s} ~(\tilde{X}-1)^{\chi_s}~,
\end{equation}
leading to (data not shown)
\begin{equation}
\Delta S_\mathsf{\,f} \sim (N-1)^{1.54(1)}~(\tilde{X}-1)^{0.975(6)}~.
\label{Eq:D_S_value}
\end{equation}
In this case, our extrapolations give ${\eta_s}(N\to \infty) \approx 1.96(1)$ and ${\chi_s}(\tilde{X} \to \infty) \approx {1.01(3)}$, suggesting the asymptotic scaling law
\begin{equation}
  \Delta S_\mathsf{\,f} \sim 
N^{2}\,\tilde{X} \sim N \, \Delta \, \mathbf{s}.
\label{eq:ds-mean-mono}
\end{equation} 
Thus, the final span is $\sim\!N$ times its minimum value.

There are several key asymptotic scaling results here. First, the \textit{diffusion ratio} scales like $\langle \Delta x_{\textsc{cm}}^2(\tau_\mathsf{\,f}) \rangle/\tau_\mathsf{\,f} \sim 1/M$, as one would expect. Second, the \textit{relative displacement ratio} $\langle \Delta x_{\textsc{cm}}^2(\tau_\mathsf{\,f}) \rangle/\Delta S_\mathsf{\,f}^2 \sim N^{-2}$ vanishes when $N \gg 1$, indicating that \textsc{cm} diffusion is not relevant. Third, the \textit{spreading velocity} $v_s \equiv \Delta S_\mathsf{\,f}/\tau_\mathsf{\,f} \sim N\,D(M)/\Delta \, \mathbf{s} \sim 1/M \tilde{X}$ is entirely controlled by the parameter $\tilde{X}$. Finally, the key parameter $\Delta \, \mathbf{s} $, which is determined by the experimental conditions $N$ and $\tilde{X}$, controls the spreading time $\tau_\mathsf{\,f}$ for a given molecular size $M$.


\begin{figure}[!ht]
\centering
\includegraphics[width = 0.482\textwidth]{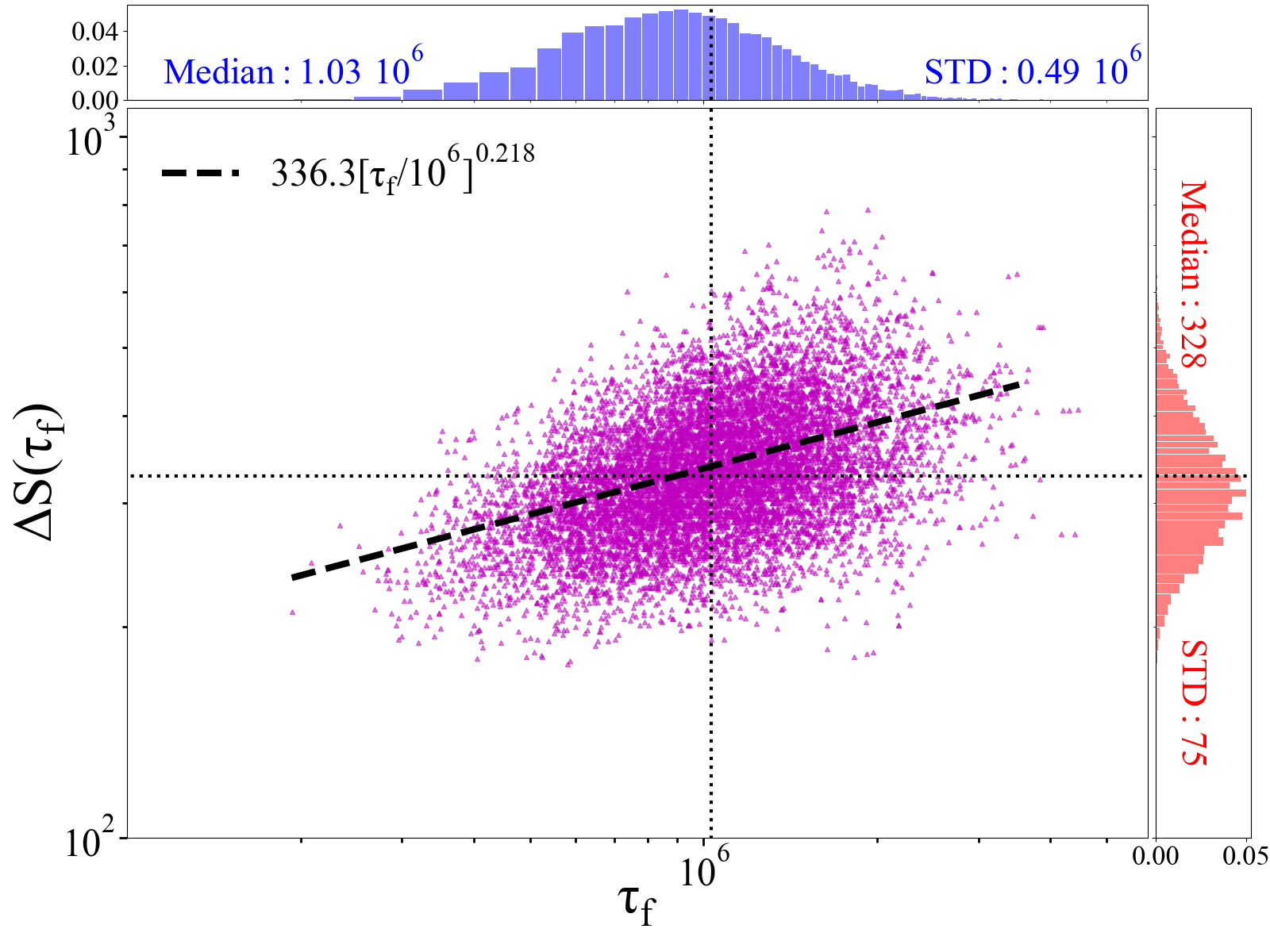}
\caption{The data points $[\tau_\mathsf{\,f},\Delta S_\mathsf{\,f}]$ for $20~000$ simulations with $N=11$ molecular fragments of size $m=10$ and a final minimum gap of $\tilde{X}=15$. The histograms on top and right show the distributions, with the median (also shown as thin dotted lines) and standard deviations. The dashed line is a power-law fit. }
\label{fig:correlation}
\end{figure}

The distribution of $[\tau_\mathsf{\,f},\Delta S_\mathsf{\,f}]$ data points, shown in Fig.~\ref{fig:correlation} for 20\,000 different simulations with $N=11$ fragments of size $m=10$ and a final gap size of $\tilde{X}=15$, demonstrates that there is wide variety of cases -- looking at mean values is not sufficient. For example, we have long times but small final spans in the bottom right quadrant, while the points in the top left quadrant correspond to rapid spreading accompanied by large spans. The power-law fit having a small slope of 0.218(1), the correlation between $\tau_\mathsf{\,f}$ and $\Delta S_\mathsf{\,f}$ is rather weak. 

It is possible to estimate the missing asymptotic coefficients in Eqs.~\ref{eq:taufdM} and \ref{eq:ds-mean-mono} using the fact that $\langle \tau_\mathsf{\,f} \rangle \approx 10^6$ and $\Delta S_\mathsf{\,f} \approx 300$ in Fig.~\ref{fig:correlation}. Since $M=mN=10\times 11=110$ while $\Delta \, \mathbf{s} =(N-1)(\tilde{X}-1)=14 \times 10 = 140$, we obtain 
\begin{equation}
 \begin{split}
    \langle \tau_\mathsf{\,f} \rangle &\approx \tfrac{1}{2}~M~\left[\Delta \, \mathbf{s}\right]^2 \\ \Delta S_\mathsf{\,f} &\approx \tfrac{1}{5}~N~~\,\Delta \, \mathbf{s}  
 \end{split}
 \label{eq:Delta_s}
\end{equation}
Although this example is not in the asymptotic limit, these approximate relations will be useful guides later.

\subsection{Bimodal distributions of molecules}

\label{sec:bimodal}

We now study a bimodal system with $N_1$ fragments of size $m_1$ and $N_2$ fragments of size $m_2$, for a total molecular size $M=N_1 m_1+N_2 m_2$. As the number of ways to distribute $N=N_1+N_2$ molecules along the tube is very large, we only study a few cases, including the simple one where the $N_1\!-\!1$ pairs of type 1 molecules are separated by $P$ type 2 molecules each, with the remaining $\delta N $ type 2 molecules placed at the end(s).  

\begin{figure}[!ht]
\centering
\includegraphics[width = 0.482\textwidth]{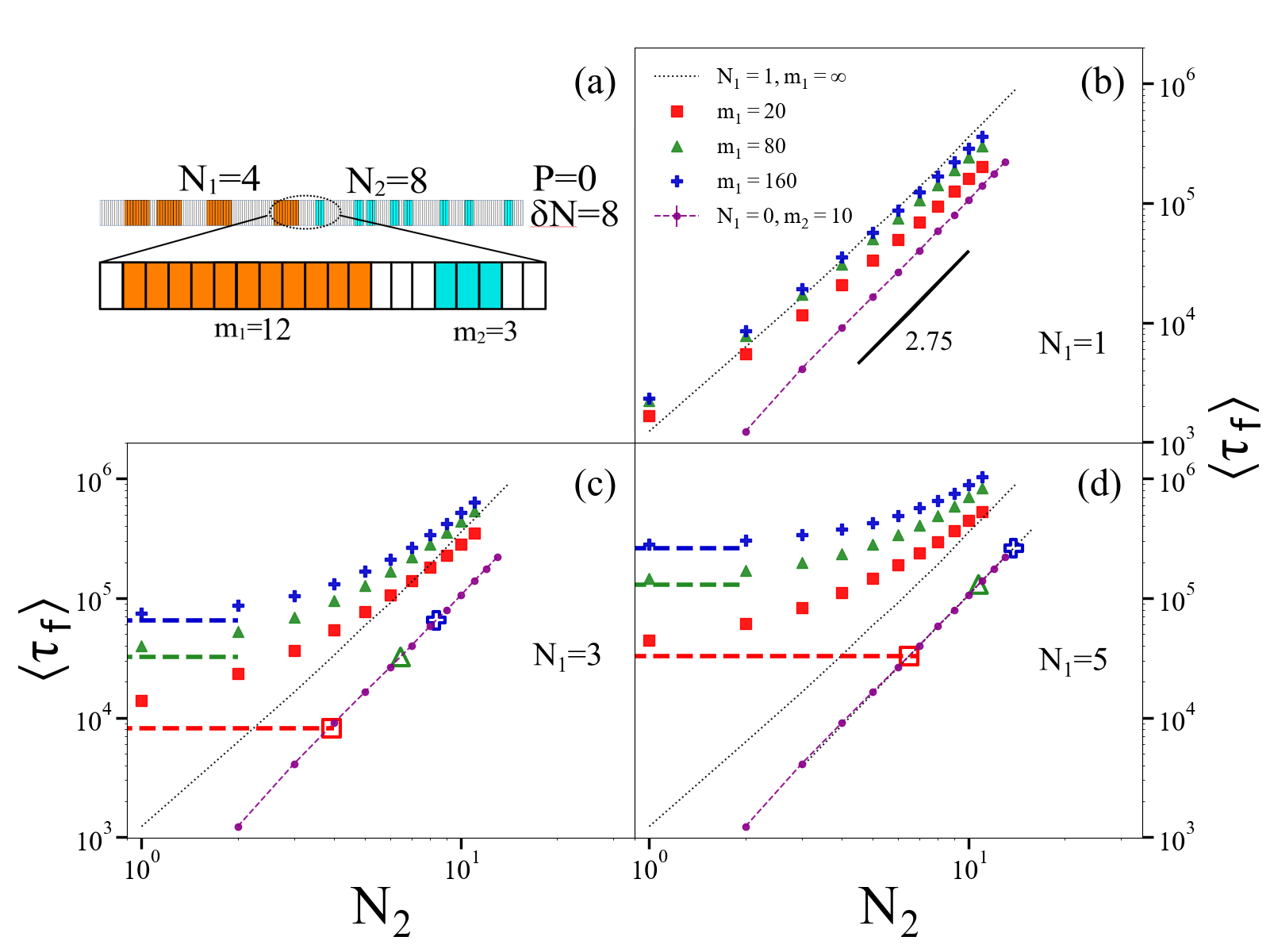}
\caption{(a) A schematic view of a segregated bimodal system with $N_1=4$ fragments of size $m_1=12$ and $N_2=8$ fragments of size $m_2=3$. The next three panels show the mean spreading time $\langle \tau_\mathsf{\,f} \rangle$ \textit{vs} the number $N_2$ of type 2 molecules of size $m_2=10$ for three different values of $m_1$; the final minimum spacing is $\tilde{X}=5$ and the ensemble size is 12\,000. In all three cases, the lowest data set is for $N_1=0$, the dashed horizontal lines correspond to $N_2=0$, while the dotted lines present the $[N_1=1;~ m_1 \to \infty ]$ limit. (b) $N_1=1$; there is no horizontal line since $\tau_\mathsf{\,f}$ is not defined when $N_2=0$ and $N_1=1$. (c) $N_1=3$. (d) $N_1=5$.  The hollow markers along the lower ($N_1=0$) line show the transition points $N_2^{*}(m_1)$ between the two asymptotic regimes. }
\label{fig:p=0_tau}
\end{figure}

An example with $M\!=\!72$, $m_1\!=\!12$, $m_2\!=\!3$, $N_1\!=\!4$, $N_2\!=\!8$ and $P\!=\!0$ is shown in Fig.~\ref{fig:p=0_tau}~(a); as $P=0$, the two types of molecules form separate groups. We show $\left<\tau_\mathsf{\,f}\right>$ vs $N_2$ data for such segregated systems in panels (b--d) for $N_1=1$, 3 and 5, respectively; we keep $m_2=10$ and $\tilde{X}=5$ fixed, and examine three values of $m_1 > m_2$ (note that $M$ is not kept fixed here). To better understand the results, we also show data for three limit cases: 1) the lowest line is for $N_1=0$; 2) the horizontal dashed lines are for $N_2=0$; 3) the dotted line is for $[N_1=1; m_1 \to \infty]$ (a single type-1 molecule acts as a fixed wall on the left).

As shown in Figs.~\ref{fig:p=0_tau}~(c--d), the spreading time plateaus at a value $\left< \tau_\mathsf{\,f}(N_1) \right>$ when $N_2\!<\!N_1$ because the $N_1$ large molecules then limit the spreading process. Conversely, when $N_2 \gg N_1$, all data sets must converge to the lowest line $\left< \tau_\mathsf{\,f}(N_2) \right>$, which corresponds to having only type-2 molecules. Although this convergence is visible for $m_1=20$ in Fig.~\ref{fig:p=0_tau}~(b), it is much slower in the other cases. Interestingly, the data sets first converge towards the dotted line, then cross it before converging towards the lowest line. This intermediate regime corresponds to a small group of $N_1 \ll N_2$ type-1 molecules of size $m_1 \gg m_2$ acting as an immobile wall from which the $N_2$ smaller molecules must move away. It is only when the relaxation time of the $N_1$ molecules has become negligible compared to that of the $N_2$ molecules that the data sets can converge towards the lowest line. We can also see this double-convergence process in Fig.~\ref{fig:p=0_tau}~(b), but there is no plateau when $N_2 \to 0$ since we then have a single (large) molecule. The transition between the two asymptotic regimes can be defined as the point where $\left<\tau_\mathsf{\,f}(N_1)\right>=\left<\tau_\mathsf{\,f}(N_2)\right>$, as shown in Figs.~\ref{fig:p=0_tau}~(c--d); Eq.~\ref{eq:tauf_scaling} predicts that this happens for a critical population size
\begin{equation}
    N_2^*=N_1 \times \left( \frac{m_1}{m_2} \right)^{{1}/{\eta_\tau}}~.
\end{equation}
This rough approximation, which works well for our data here with $\eta_\tau  \approx 2.75$ (not shown), indicates that the transition point is a rather weak function of the size ratio $m_1/m_2$, especially given the asymptotic exponent $\eta_\tau =3$.

Figure~\ref{fig:p=0_tau} showed data for segregated bimodal systems ($P=0$). The opposite limit is when $N_2=(N_1-1) \times P$: all of the type-2 molecules are then located between the larger type-1 molecules. Given Eq.~\ref{Eq:tau_f_value}, it is not surprising that we recover a scaling form $\tau_\mathsf{\,f}(P \gg 1) \sim P^{\eta_\tau}$ (data not shown) in the limit where the type-2 molecules dominate.

As described previously, the total mass $M$ is fixed when the molecule is cut into $N$ pieces in an experimental situation. We thus finish this section by looking at simple bimodal systems with a fixed $M$ but varying mass ratios $\mu\!\equiv\!{m_2}/{m_1}$. The questions here include: 1) Is $\tau_\mathsf{\,f}$ smaller in the presence of several small molecules or a few large ones? 2) Does having large molecules at both ends of the sequence slow down spreading? 3) What is the effect of the mass asymmetry $\mu$?

\begin{figure}[!ht]
\centering
\includegraphics[width = 0.482\textwidth]{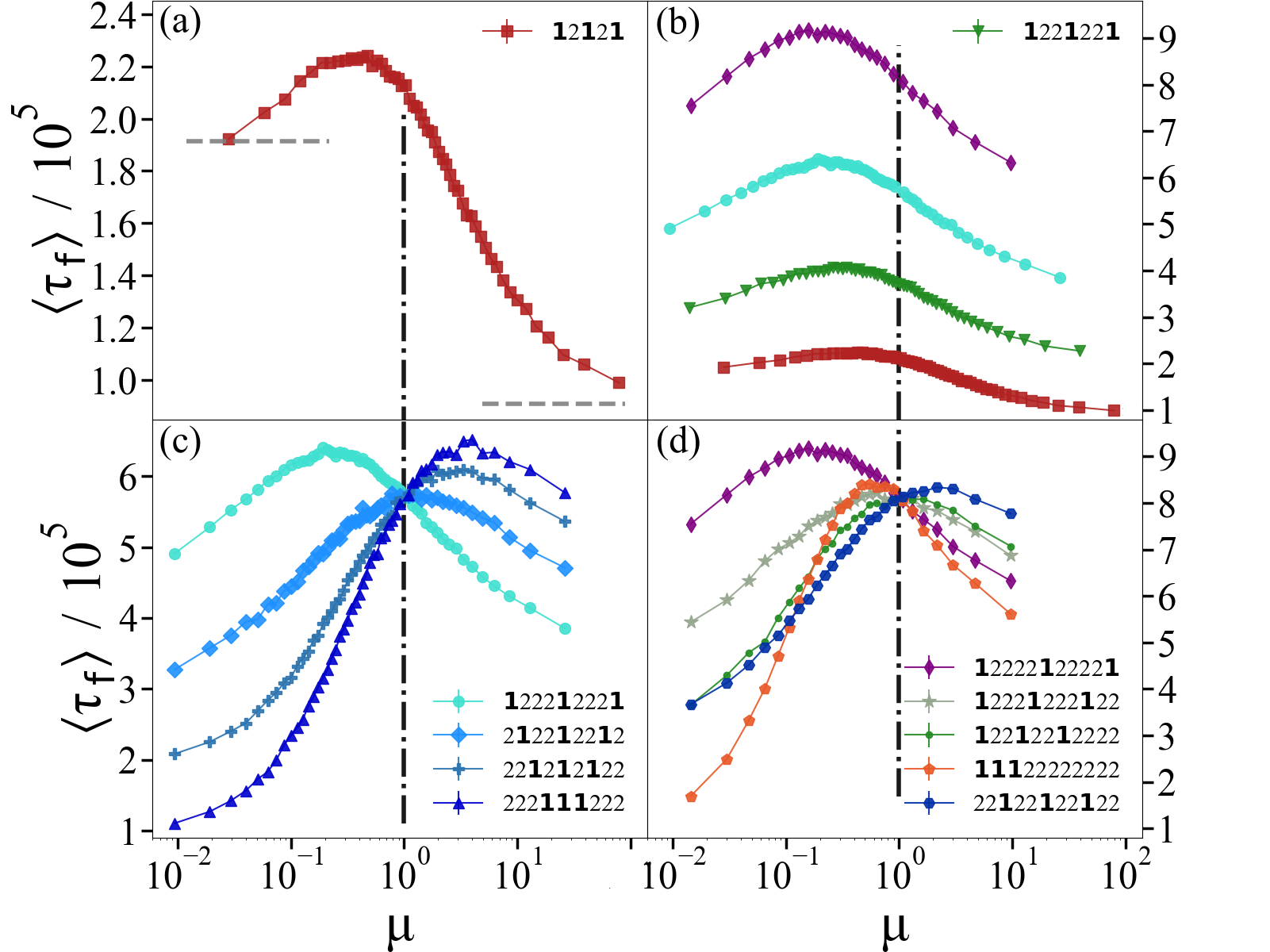}
\caption{Mean spreading time $\langle \tau_\mathsf{\,f} \rangle$ \textit{vs} mass ratio $\mu=m_2/m_1$. The systems are described symbolically as series of \textbf{1} and 2. The total molecular weight is fixed at $M=648$, while $\tilde{X} = 5$ and $N_1=3$. (a) $N_2=2$ and $P=1$; (b) $N_2=2P$ with, from top to bottom, $P=4 ({\color{Purple}\blacklozenge}),3 ({\color{cyan}\LARGE \bullet}),2 ({\color{ForestGreen}\blacktriangledown}),1 ({\color{BrickRed}\blacksquare})$. The $P=2$ sequence is shown. (c) Four systems with $N_2=6$ are compared: the type-1 molecules are systematically moved towards the centre of the $P=3$ sequence of panel B; (d) Five systems with $N_2=8$ are compared.}
\label{fig:mu_inc}
\end{figure}

Figure~\ref{fig:mu_inc}~(a) shows $\langle \tau_\mathsf{\,f} \rangle$ \textit{vs.} the mass ratio $\mu$ for a system where $N_1=3$, $N_2=2$, $M=648$ and $P=1$ are kept fixed (this sequence thus reads \textbf{1}2\textbf{1}2\textbf{1}). Interestingly, the mean spreading time $\langle \tau_\mathsf{\,f}(\mu) \rangle$ is not a monotonic function; instead, it peaks at about $\mu=0.45(3)$ and plateaus in both limits (the two horizontal dashed lines mark the limits
when the smallest molecules are
of size $m=1$). Since $\left<\tau_\mathsf{\,f}\right> \sim m (N-1)^{\eta_\tau}$ with $\eta_\tau > 2$ according to Eq.~\ref{Eq:tau_f_value}, we do expect molecular spreading to take more time with three large molecules of size $m_1=215$ rather than two large molecules of size $m_2=322$. Moreover, as we shall see, large molecules at (or near) the ends have a greater impact than those in the middle. Equation~\ref{eq:Delta_s} predicts a mean value of $\langle \tau_\mathsf{\,f} \rangle \approx  0.83~10^5$, close to the lowest plateau; unsurprisingly, the dependence of the mean spreading time $\langle \tau_\mathsf{\,f} \rangle$ on the mass ratio $\mu$ is not captured by Eq.~\ref{eq:Delta_s}.

Data for similar systems but different $P$ values are shown in Fig.~\ref{fig:mu_inc}~(b): these sequences read $\textbf{1}[2]_\mathrm{P}\textbf{1}[2]_\mathrm{P}\textbf{1}$ (the $P=2$ case is shown in the top right corner). The non-monotonic curves move up and drift towards smaller values of $\mu$ when $P$ increases. These results clearly show the effect of having more, albeit smaller, molecules in the system. 

In Fig.~\ref{fig:mu_inc}~(c), we start with the $P=3$ case and systematically move the type-1 molecules towards the centre while preserving the left-right symmetry. Clearly, having large molecules near the ends increases the spreading time. In this case, Eq.~\ref{eq:Delta_s} predicts $\langle \tau_\mathsf{\,f} \rangle \approx 3.3~10^5$, an excellent estimate overall.

Finally, in Fig.~\ref{fig:mu_inc}~(d) we start with the $P=4$ case and move the type-1 molecules towards one of the ends. Segregating the larger molecules generally decreases the spreading time, mostly because the smaller molecules can then move more freely. We also show data for the sequence $22\textbf{1}22\textbf{1}22\textbf{1}22$: this curve converges with that of the $\textbf{1}22\textbf{1}22\textbf{1}2222$ sequence in the $\mu \ll 1$ limit. These two sequences share the same $\textbf{1}22\textbf{1}22\textbf{1}$ core; when $\mu \ll 1$, the type-2 molecules outside this core play a minor role. Equation~\ref{eq:Delta_s} predicts $\langle \tau_\mathsf{\,f} \rangle \approx 5.2~10^5$ for panel D, again an excellent estimate overall.

We thus conclude that having a large number of small molecules generally leads to a longer spreading process, while large molecules placed at the end tend to act as slow-moving walls. We also notice that for a given sequence, the spreading time is a non-monotonic function of the size ratio $\mu$. All of these results suggest that the precise location of the molecules in a random sequence $(m_1,m_2,..,m_N)$ can affect the mean spreading time, an effect that Eq.~\ref{eq:Delta_s} does cannot capture.

\subsection{Random distributions: main results}
\label{subsec:main_random}

We now study the central problem of this article, i.e., that of a molecule of size $M=\sum_{i=1}^{N} m_i$ randomly cut into $N$ pieces of size $m_i$ inside a narrow channel. This process generates an ordered sequence of molecular sizes $(m_i)$ along the channel. Since we assumed that $\tilde{X}$ is the minimum gap that can be measured, we also use $\tilde{X}$ as the minimum molecular size that can be visualized; therefore, all $m_i$'s $\in [\tilde{X},M-(N-1)\tilde{X}]$. 

We first address the question of the mean spreading time for random sequences. Our data (not shown) indicate that Eq.~\ref{eq:tauf_scaling} now reads (R stands for Random):
\begin{equation}
\begin{aligned}
 \langle \tau_\mathsf{\,f} \rangle & ~\sim~ \overline{m} ~(N-1)^{\eta^R_\tau} ~~~~~~(\tilde{X}-1)^{\chi^R_\tau}~, \\ 
 & ~\sim~ \overline{m} ~(N-1)^{2.60(1)} ~(\tilde{X}-1)^{1.70(5)}~.
\end{aligned}
\label{eq:tauf_scaling_rand}
\end{equation}
Extrapolating our results, we find ${\eta^R_\tau}(N\to \infty) \approx 2.91(3)$ and ${\chi^R_\tau}(\tilde{X} \to \infty) \approx 1.97(3)$, suggesting the asymptotic behavior
\begin{equation}
   \langle \tau_\mathsf{\,f} \rangle ~\sim~ \overline{m}~(N-1)^3 (\tilde{X}-1)^2 \sim M \times \left[\Delta \, \mathbf{s}\right]^2, 
\end{equation}
identical to Eq.~\ref{eq:taufdM}, the relation for monodisperse systems, except that $m \to \overline{m}$. Similarly, 
\begin{equation}
\begin{aligned}
 \Delta S_\mathsf{\,f} & \sim ~(N-1)^{\eta^R_s} ~~~~~~(\tilde{X}-1)^{\chi^R_s}~, \\ 
 & \sim ~(N-1)^{1.52(3)} 
 ~(\tilde{X}-1)^{0.88(1)}~,
\end{aligned}
\label{eq:span_scaling_rand}
\end{equation}
with the extrapolated values ${\eta^R_s}(N\to \infty) \approx 2.0(1)$ and ${\chi^R_s}(\tilde{X} \to \infty) \approx 0.97(3)$. We thus conclude that Eq.~\ref{eq:ds-mean-mono}, for $\Delta S_\mathsf{\,f} \sim N \, \Delta \, \mathbf{s} $, is also unaffected by randomness. Even though random distributions behave like monodisperse ones on average, it does not mean that the average behavior is a good description of specific cases (this was also true in the previous section for bimodal mass distributions).

Three questions come to mind here. First, given that the distribution of spreading times is broad even for monodisperse systems (e.g., see Fig.~\ref{fig:correlation}), does the random cutting process add further variability to $\tau_\mathsf{\,f}$? Second, what are the differences between the most likely random sequences $(m_i)$ and the extreme ones -- e.g., the monodisperse ones? Finally, since molecular order matters, what is the impact of simply permuting elements in a sequence $(m_i)$?

\begin{figure}[!ht]
\centering
\includegraphics[width = 0.46\textwidth]{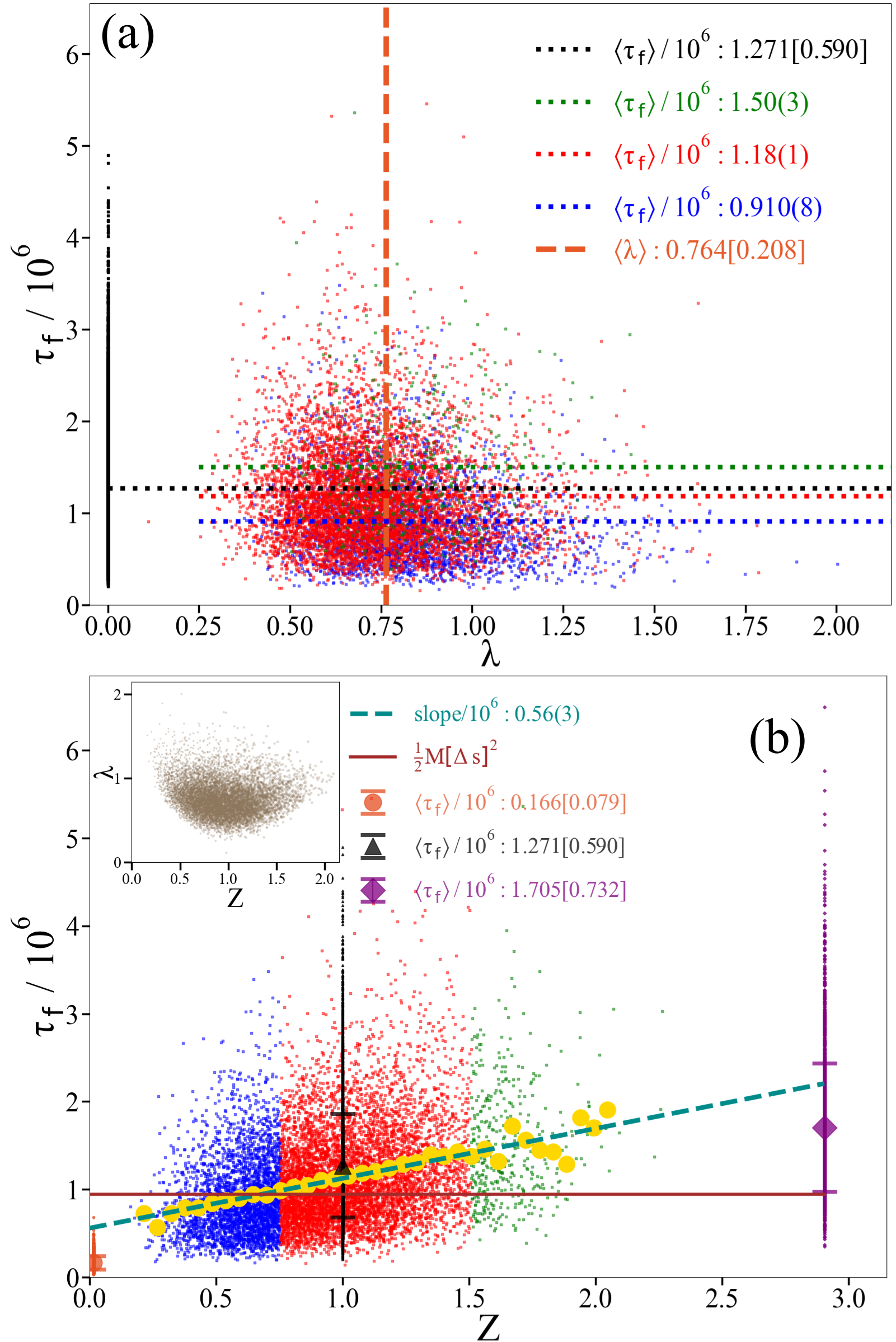}
\caption{(a) Spreading time $\tau_\mathsf{\,f}$ \textit{vs} $\lambda$ for $10^4$ molecules of size $M\!=\!600$ randomly cut into $N\!=\!9$ pieces of minimum size $\tilde{X}\!=\!8$. The means and standard-deviations (in $[\cdot]$) are $\langle \tau_\mathsf{\,f} \rangle=1.110~[0.563]~10^6$ and $\overline{\lambda}=0.764~[0.208]$. The data for the monodisperse system (in grey at $\lambda=0$) has a mean of $\langle \tau_\mathsf{\,f} \rangle=1.271~[0.590]~10^6$. The points are colored based on their $Z$ values, as defined in Eq.~\ref{eq:Z}. The maximum value of $Z$ being $Z_m = 2.26$ here
, the color code is as follows: green for $\frac{2}{3}\,Z_m \le Z \le Z_m$; red for $\frac{1}{3}\,Z_m \le Z < \frac{2}{3}\,Z_m$; and blue for $0 \le Z < \frac{1}{3}\,Z_m$. (b) Spreading time $\tau_\mathsf{\,f}$ \textit{vs} $Z$. The cyan dashed line (\textcolor{cyan}{\textbf{--\,--\,--}}) is a linear fit to the binned data (large golden dots \textcolor{Dandelion}{\Large $\bullet$}). The three vertical distributions (2,000 data points each) correspond to $Z=0.014$ (the theoretical minimum $Z$ value), $Z=1$ (the monodisperse case) and $Z=2.904$ (the theoretical maximum $Z$ value). Insert: $\lambda$ \textit{vs} $Z$.}
\label{fig:random_1}
\end{figure}

The randomness of a sequence $(m_i)$ can be described by the ratio $\lambda= {\mathrm{STD}[(m_i)]}/\overline{m}$ between the standard-deviation (STD) and the mean $\overline{m}=M/N$ of its $N$ elements. Figure~\ref{fig:random_1}~(a) presents $\tau_\mathsf{\,f}$ \textit{vs} $\lambda$ data for $10^4$ molecules of size $M=600$ randomly cut into $N=9$ pieces with $\tilde{X}=8$. The mean values are $\overline{\lambda}=0.764(2)$ and $\langle \tau_\mathsf{\,f} \rangle=1.110(6)~10^6$, while Eq.~\ref{eq:Delta_s} predicts $\langle \tau_\mathsf{\,f} \rangle \approx 0.94~10^6$, consistent with our data. The weak correlation between $\tau_\mathsf{\,f}$ and $\lambda$ decreases rapidly when $N$ increases (data not shown). The data points on the left (in grey) correspond to the monodisperse case $\lambda=0$; we then have $\langle \tau_\mathsf{\,f} \rangle=1.271(5)~10^6$, larger than the mean value for random cuts.

The distribution of $\lambda$ values narrows as $N$ increases, while $\overline{\lambda}$ increases toward an asymptotic value (data not shown).  Moreover, $\lambda$ does not provide information on the location of the largest molecules in an ordered sequence $(m_i)$, which is an important factor, as shown in previous sections. Thus, we need another parameter to characterize $(m_i)$, a parameter that measures how much mass is located near the two ends. We propose to use (defined here for $N$ odd):
\begin{equation}
    Z ~ = ~  \frac{\left[\sum\limits_{i=1}^{n}(n-i)\,m_i\right] \times \left[\sum\limits_{i=n}^{N}(i-n)\,m_i\right]}{\left[\tfrac{M/N}{8}~(N^2-1)\right]^2}
\label{eq:Z}
\end{equation}
where $n=\frac{1}{2}(N+1)$ marks the central molecule of the sequence. The two terms in the numerator are basically torque-like quantities from the two halves of the sequence. The denominator, which is used to normalize the Z parameter, is the value of the numerator for a monodisperse system. Larger values of $Z$ indicate the presence of large molecules near both ends (indices $i=1,N$), while a small $Z$ means that most large molecules are located near the center (index $i=n$).

We colored the data points in Fig.~\ref{fig:random_1}~(a) based on their $Z$ values; the horizontal dashed lines give the mean spreading times $\langle \tau_\mathsf{\,f} \rangle$ for each of the three colors. We note that $\langle \tau_\mathsf{\,f} \rangle$ increases from blue to green, and that the $Z=1$ monodisperse case is in the middle. Although these results confirm that having large molecules near \textit{both} ends slow down spreading, there is substantial overlap between the three groups. 

Figure~\ref{fig:random_1}~(b) shows that $\tau_\mathsf{\,f}$ tends to increase with $Z$, confirming that the location of the largest molecules matters. We also note that the monodisperse system is located in the red region, at $Z=1$, and that the binned data can be fitted with a straight line. We added data sets for two extreme cases. The sequence  (\textbf{272} $[8]_7$ \textbf{272}), which has the maximum possible value of $Z=2.904$, has a mean time $\langle \tau_\mathsf{\,f} \rangle = 1.705 ~ 10^6$, while the sequence $([8]_4 \textbf{536} [8]_4)$, with the minimum possible value of $Z=0.014$, gives $\langle \tau_\mathsf{\,f} \rangle = 0.166~10^6$. In both cases, the STD of the distribution of spreading times is $\approx 45\%$ of its mean (this is also true for $Z=1$). The fit does not extrapolate to these two limits, pointing to a change in behavior for extreme cases; however, it nearly captures the monodisperse case.

The inset of Fig.~\ref{fig:random_1}~(b) shows that there is an overall nonmonotonic relationship between the randomness parameters $\lambda$ and $Z$. This is not surprising since, for any given value of $\lambda$, it is possible to reduce (increase) the value of $Z$ by moving the larger molecules toward the center (ends) of the sequence.

The main conclusion comes directly from Fig.~\ref{fig:random_1}~(b): while the mean spreading time increases from about $0.5~10^6$ to $2~10^6$ over the full range of $Z$ values, the standard deviations are of order $10^6$. In other words, the correlation between the spreading times and the details of the $(m_i)$ sequence is roughly comparable to the randomness of the Brownian spreading process itself. 

\subsection{\label{subsec:permutation}Random distributions: permutations}

\begin{figure}[!ht]
\centering
\includegraphics[width = 0.482\textwidth]{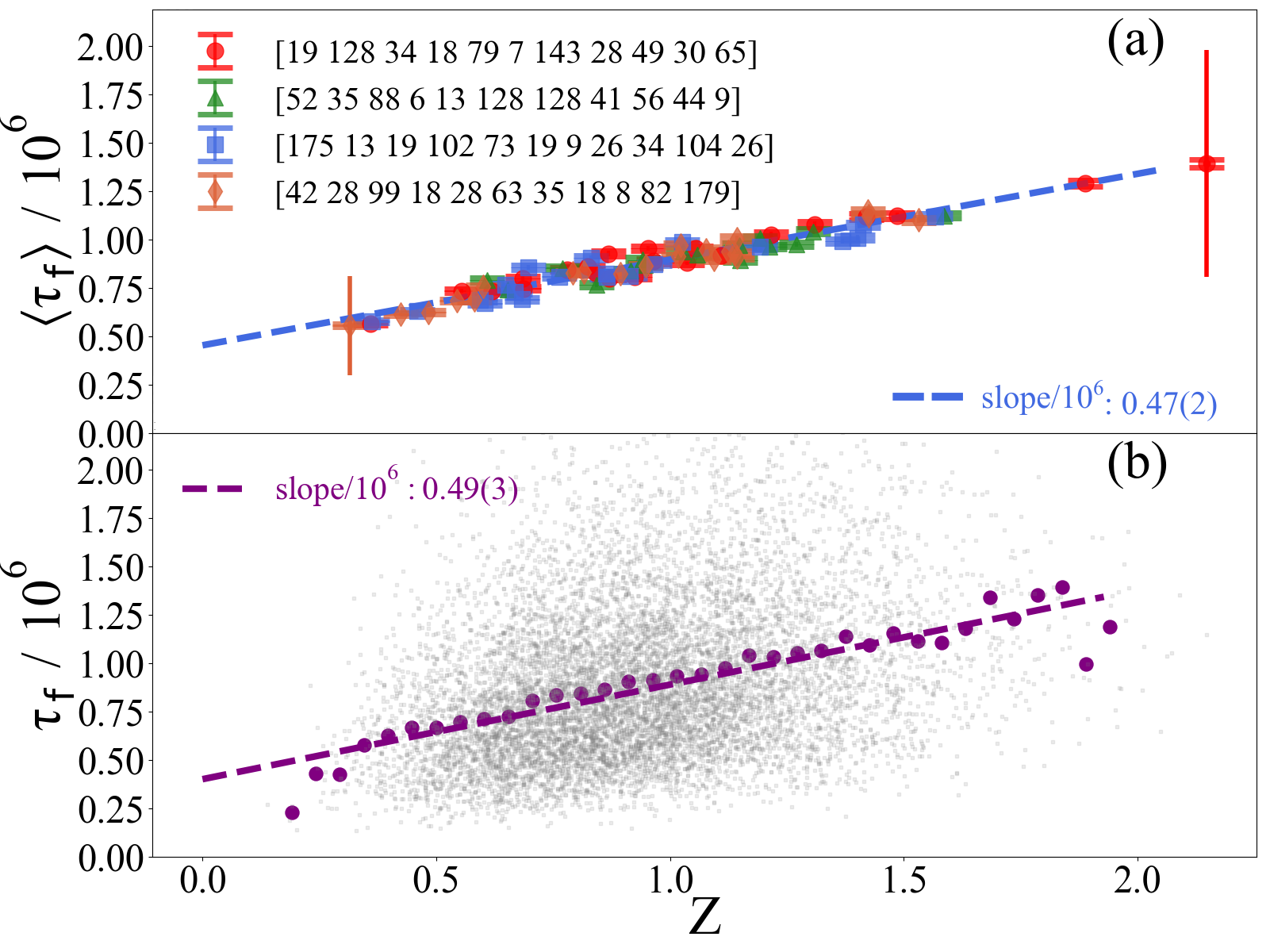}
\caption{(a) Mean spreading time $\langle \tau_\mathsf{\,f} \rangle$ vs $Z$ for systems of size $M\!=\!600$ cut into $N\!=\!11$ pieces of minimum size $\tilde{X}\!=\!6$. Four random sequences with the following $\lambda$ values are presented: 0.792 (\textcolor{red}{red $\bullet$}), 0.757 (\textcolor{ForestGreen}{green $\blacktriangle$}), 0.924 (\textcolor{blue}{blue $\blacksquare$}), and 0.875 (\textcolor{orange}{orange $\blacklozenge$}). Each one was shuffled 20 times -- thus generating sequences with 20 different values of $Z$. Each data point is an average over 1000 simulations; the error bars are comparable to the size of the symbols (the vertical lines mark the standard-deviations for the two end points). (b) Spreading time $\tau_\mathsf{\,f}$ vs $Z$ for $10\,000$ random systems (no permutations) with the same parameters as in (a). The dashed line is a linear fit to the binned data (large dots).}
\label{fig:permutation}
\end{figure}

Although shuffling the elements in a sequence $(m_i)$ does not affect $\lambda$, it does change the value of our $Z$ parameter. In this section, we examine how such permutations affect the spreading time.

There are $\Omega=\nicefrac{N!}{2}$ ways of placing $N$ different objects on a line. For example, there are $\Omega \approx 20$~million possible permutations for a given molecular sequence $(m_i)$ containing $N=11$ molecules ($\Omega$ is smaller if some molecules happen to have the same size). As $\Omega$ is very large, we use the following sampling process. We first generate a random sequence $\Lambda=(m_i)$ for a given total molecular weight $M$, number of fragments $N$, and minimum molecular size $\tilde{X}$. We then choose a random subset of $\Omega_\Lambda^*=20$ permutations
. The latter are simulated 1000 times each in order to obtain their mean spreading times. Finally, the process is repeated for different initial sequences $\Lambda$.

Figure~\ref{fig:permutation}~(a) shows $\langle \tau_\mathsf{\,f} \rangle$ \textit{vs} $Z$ data for four initial sequences. The linear increase observed in Fig.~\ref{fig:random_1}~(b) is recovered in all cases. In fact, these four systems give essentially identical data (except that the largest and smallest values of $Z$ strongly depend on the initial sequence). The four data sets were fitted together to obtain a slope of $0.47(2) \, 10^6$. Figure~\ref{fig:permutation}~(b) shows the data for 10~000 random systems (no permutations), similar to what we presented in Fig.~\ref{fig:random_1}~(b); the linear fit gives a slope of $0.49(3) \, 10^6$, indicating that the relation between $Z$ and the spreading time is universal. 

Our data, for both random \textit{and} shuffled sequences, show that Brownian noise is as important as the details of the molecular sequence. In fact, the latter only shifts, usually rather slightly, the center of a broad distribution of spreading times.

\subsection{\label{subsec:extreme}Random Distributions: Extreme Cases}

\begin{figure}[!ht]
\centering
\includegraphics[width = 0.482\textwidth]{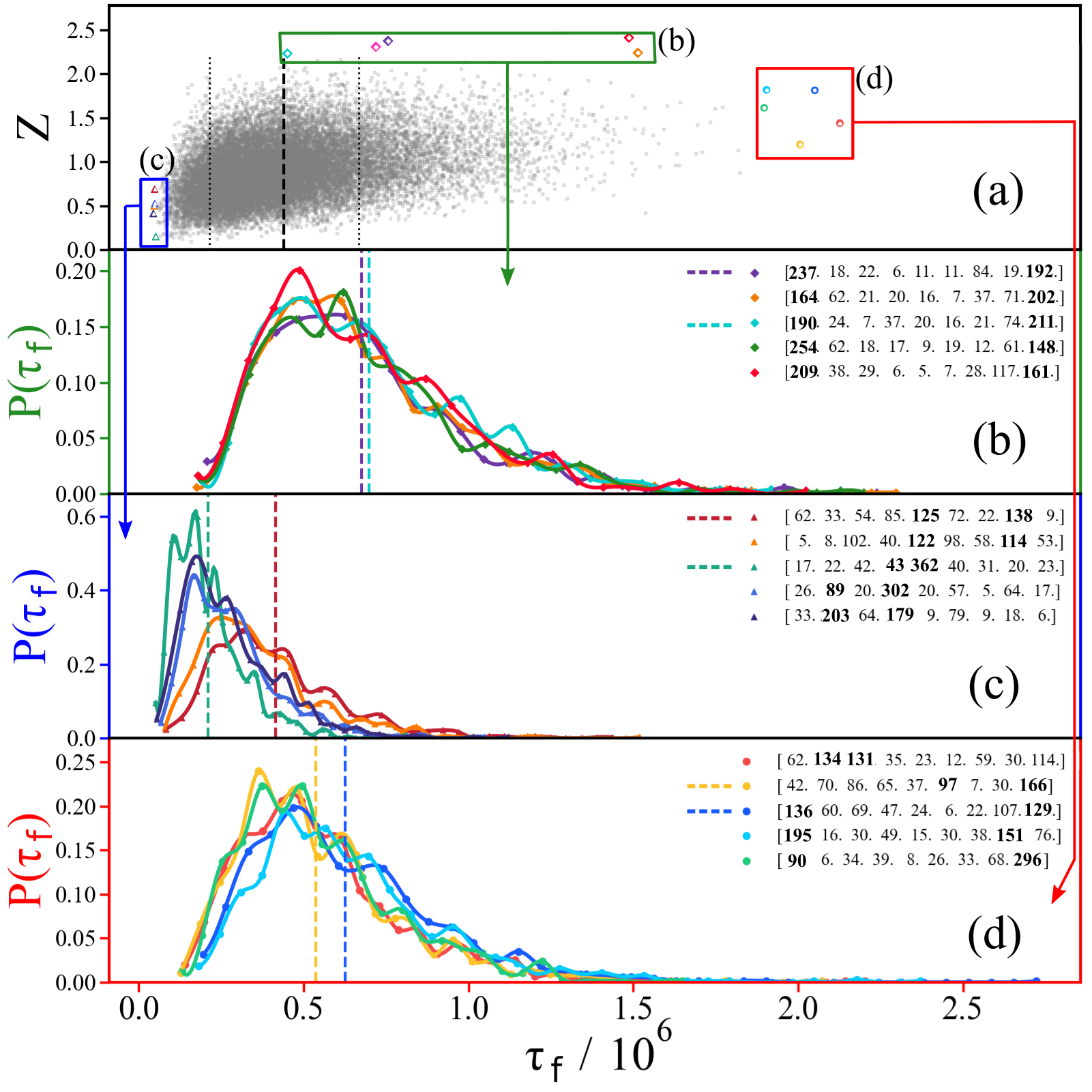}
\caption{Data for a molecule of size $M=600$ randomly cut into $N=9$ pieces of minimum size $\tilde{X}=5$. (a) Spreading time $\tau_\mathsf{\,f}$ \textit{vs} $Z$ for 10~000 random systems; the dashed lines give the mean and width (STD) of the distribution of points. (b) The distribution function of spreading times for the 5 sequences with the largest values of $Z$ in (a). (c-d) The distribution function of spreading times for the 5 sequences with the shortest (largest) values of $\tau_\mathsf{\,f}$ in (a). In panels (b-d), the dashed lines give the maximum and minimum values of the means of the five distributions shown. The chosen sequences are given in the legend, with the two largest molecules in bold. }
\label{fig:extream}
\end{figure}

To better understand our results, we now focus on extreme cases, specifically the outliers in Fig.~\ref{fig:extream}~(a). A total of 15 cases were selected for further analysis, and each case was simulated 2000 times. The distribution of spreading times $\tau_\mathsf{\,f}$ for the 5 cases with the shortest (longest) times is shown in panel (c) (panel (d)), while panel (b) shows the distributions for the 5 cases with the largest values of $Z$.

The distribution functions in Fig.~\ref{fig:extream}~(d) are similar to each other; unsurprisingly, their means are larger than the global mean separation time shown in Fig.~\ref{fig:extream}~(a). All but one of these sequences have large molecules near their ends; the exception has a molecule of size 97 in the middle and possesses the shortest separation time in this group. All five of these outliers are found far in the high-$\tau_\mathsf{\,f}$ tail of their distributions.

Figure~\ref{fig:extream}~(c) shows that the distributions do not overlap for the low-$\tau_\mathsf{\,f}$ outliers, all of which have large objects near their midpoints. The shortest times are found for the smallest $Z$ value (in green): the central molecule (size 362) is over 50\% of the total weight. The largest value of $Z$ (in red) corresponds to the longest times: this case has a large molecule in the middle, but the largest one is close to an end. All 5 of these outliers are found far in the low-$\tau_\mathsf{\,f}$ tail of their distributions.

As Fig.~\ref{fig:extream}~(a) shows, the $Z$ values of the high-$\tau_\mathsf{\,f}$ outliers are larger than those of the low-$\tau_\mathsf{\,f}$ outliers. Yet, there is a substantial amount of overlap between the distribution functions found in panels (c) and (d), demonstrating that Brownian variability is at least as important as the $Z$ factor.

The distributions for the five outliers with the highest values of $Z$ are similar to each other in Fig.~\ref{fig:extream}~(b), and even share the same mean and standard deviation. All five sequences feature very large molecules at both ends, a factor we identified as responsible for slowing down the spreading process.

\begin{figure}[!ht]
\centering
\includegraphics[width = 0.482\textwidth]{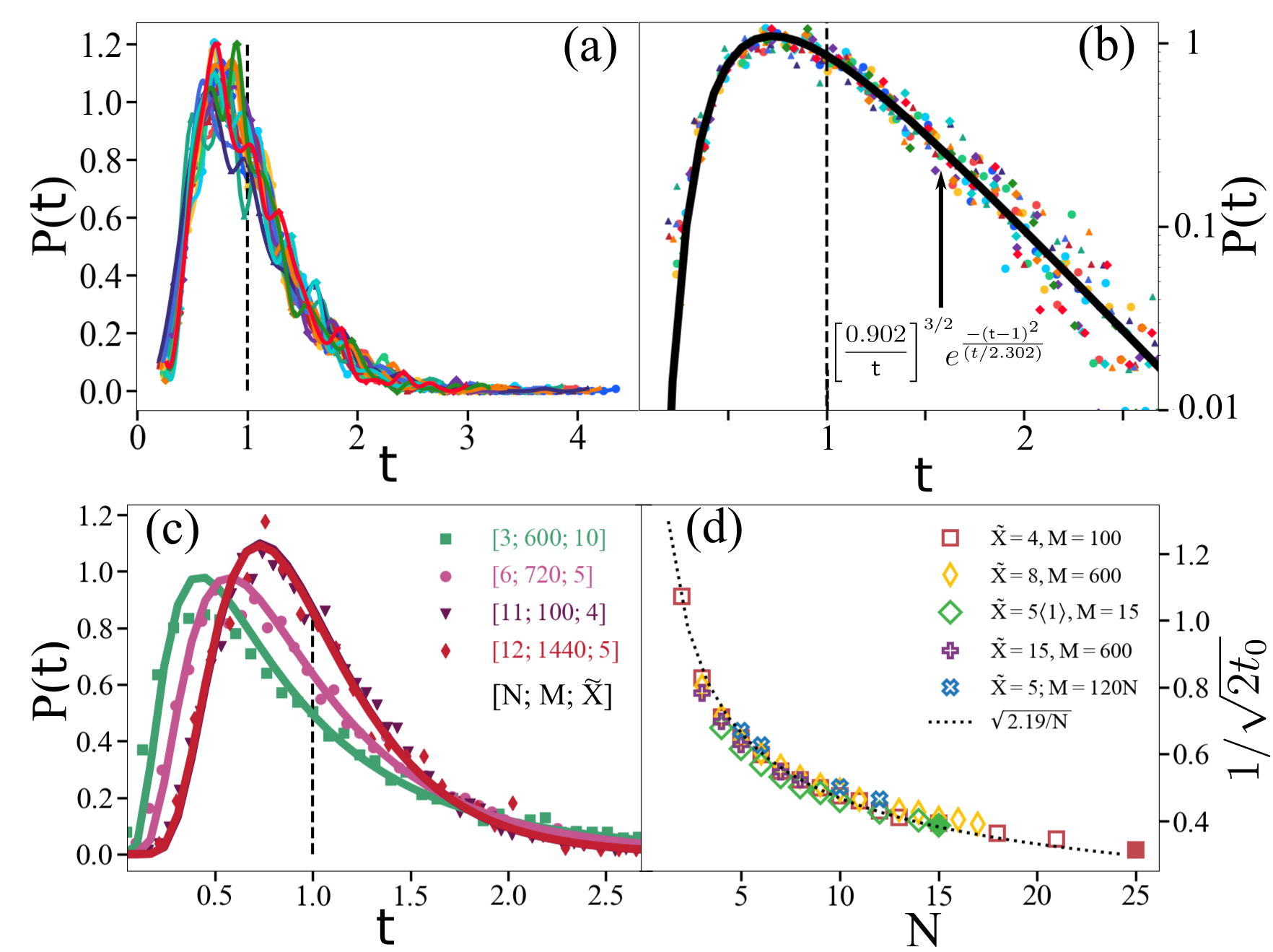}
\caption{(a) Distribution function of rescaled spreading times $\mathsf{t} \equiv \tau_\mathsf{\,f}/\langle \tau \rangle$ for the 15 outlier cases chosen in  Fig.~\ref{fig:extream}. The bin width is 0.1135. (b) Same data in a semi-log format; the solid line is the function $P(\mathsf{t}) = \left[0.902/\mathsf{t}\right]^{3/2} e^{-(\mathsf{t}-1)^2/(t/2.302)} $. (c) Data for four systems with different parameters. (d) The STD of the distributions, $1/\sqrt{2t_0}$, as a function of $N$ for 5 different systems; the filled markers correspond to the largest value of $N$ allowed (i.e., a monodisperse system), while the dashed lines show the fit $(2.19/N)^{\nicefrac{1}{2}}$. Note that for the green diamonds (\textcolor{ForestGreen}{$\lozenge$} and \textcolor{ForestGreen}{$\blacklozenge$}) we used $\tilde{X}=5$ for the minimum final gap, but $\tilde{X}=\langle 1 \rangle $ for the minimum fragment size.}
\label{fig:uni_distr}
\end{figure}

Figures~\ref{fig:extream}~(b-d) suggests that one could collapse all distribution functions by rescaling $\tau_\mathsf{\,f}$ by its mean value $\langle \tau_\mathsf{\,f} \rangle$ in each case. Figure~\ref{fig:uni_distr}~(a) shows that it is indeed so for the 15 outlier cases studied above. The standard-deviation of these rescaled curves is $0.466(7)\,[0.027]$. We thus conclude that rescaled distribution functions are not affected by the initial sequence (however, the mean values $\langle \tau_\mathsf{\,f} \rangle$ are). A good description of these curves is obtained using the following empirical normalized function, inspired by the problem of a Brownian particle's first-passage time\cite{ref:FPTDist} 
\begin{equation}
    P(\mathsf{t}) =  \frac{\sqrt{\mathsf{t_0}/\pi}}{\mathsf{t}^{3/2}}~ e^{-(\mathsf{t}-1)^2/(t/\mathsf{t_0})}~,
\label{eq:unidistime}
\end{equation}
where $\mathsf{t} \equiv \tau_\mathsf{\,f}/\langle \tau_\mathsf{\,f} \rangle$. This function has a mean of 1 and a variance of $1/2\mathsf{t_0}$. Given that the standard deviation of our data is $0.466(7)$, this predicts $\mathsf{t_0}=2.30(7)$ and a maximum at $t=0.726(7)$. As Fig.~\ref{fig:uni_distr}~(b) shows, this function is excellent over the entire range of spreading times.

Equation~\ref{eq:unidistime} provides an excellent description of the distribution functions for all the systems we have examined (four more examples are shown in panel (c) for different choices of $N$, $M$ and $\tilde{X}$). As panel (d) shows, the standard deviation of the scaled distributions (related to $t_0$ in Eq.\ref{eq:unidistime}) varies with $N$ but not with $\tilde{X}$, suggesting that Eq.~\ref{eq:unidistime} may be a convenient tool for describing this problem. The fit in Fig.~\ref{fig:uni_distr}~(d) shows that the STD decreases as $1/\sqrt{N}$, with the limit being that of a monodisperse system (see the filled markers). Interestingly, the transition from a random to a monodisperse population of fragments does not affect the validity of Eq.~\ref{eq:unidistime}. 

In this paper, we chose to have the minimum final gap and the minium fragment size equal to each other (denoted $\tilde{X}$). To demonstrate that our results do not depend on this choice, Fig.~\ref{fig:uni_distr}~(d) includes a data set (green diamonds) with $\tilde{X}=5$ as the minimum final gap, but $\tilde{X}=1$ for the minimum fragment size. The data points fall on the very same universal curve.

\section{\label{sec:conclution}Conclusion}

Mapping a long DNA molecule by cutting it in a narrow channel  is a well-defined experimental situation that can be seen as a first-passage problem involving multiple moving components coupled by the single-file diffusion constraint. We used a discretized Monte Carlo simulation algorithm to investigate the relationship between the spreading time $\tau_\mathsf{\,f}$ (defined as the time required to achieve the desired final state for the first time after the enzymes have cut the DNA) and the experimental parameters related to the molecule (specifically, the total molecular weight $M$ and the number of fragments generated by the cutting process, $N$) and the device (the limited lateral resolution $\tilde{X}$). A variety of monodisperse, bimodal, and random systems with increasing system complexity were examined.

In order to better understand how these parameters influence the spreading time $\tau_\mathsf{\,f}$, we must also consider the increased span of the system, i.e. the increase of the linear space occupied by the population of DNA fragments during the process, $\Delta S_\mathsf{\,f}$. A scaling analysis of the problem begins with the characteristic length scale $\Delta \mathbf{s} \equiv  \min[\Delta S_\mathsf{\,f}] \sim N \tilde{X}$. The mean square displacement of the center-of-mass of the population during the spreading process should scale as $\Delta x_{\textsc{cm}}^2(\tau_\mathsf{\,f}) \sim [\Delta \mathbf{s}]^2 \sim N^2 \tilde{X}^2$ giving a spreading time $\tau_\mathsf{\,f} \sim \Delta x_{\textsc{cm}}^2(\tau_\mathsf{\,f}) /D(M) \sim M [\Delta \mathbf{s}]^2 \sim M N^2 \tilde{X}^2$. The outward entropy-driven velocity of the molecules should scale as $v_s \sim D(m)/\Delta \mathbf{s} \sim 1/M \tilde{X}$ since they have an average molecular size of $m=M/N$. As a result, we expect the final span of the population to scale as $\Delta S_\mathsf{\,f} \sim v_s\,\tau_\mathsf{\,f} \sim N\, \Delta \mathbf{s} \sim N^2\, \tilde{X}$.   Our simulation data validated these scaling laws in the asymptotic limit. However, we observed strong finite-size effects for any experimentally meaningful value of $N$, as well as sequence-dependent effects.

While ensemble averages over random sequences corroborate the scaling laws presented above, we observed considerable variability in the observed spreading times (more precisely, the standard deviations are $\approx 50\%$ of the means). In order to shine light on this, we studied a toy bimodal system of fragments. The results clearly indicated that the placement of the largest molecules plays a major role in determining the expected spreading time. In particular, large molecules placed near the ends slow the spreading, while spreading tends to be rapid if most of the mass is located in the middle.

The results we obtained for bimodal systems of molecules prompted us to pose a fundamental question: To what extent does the polydispersity of a random molecular sequence $(m_i)$ affect the variability of its spreading time? Is this additional source of variability as important as Brownian motion? 

We first showed that the standard deviation of the molecular size distribution of a given sequence $(m_i)$ is a very weak predictor of its mean spreading time $\langle \tau_\mathsf{\,f} \rangle$. This outcome was to be expected, given that the value of the STD does not consider the location of the largest fragments along the channel.

We then proposed a new parameter, denoted Z, that describes the "weight torque" of the molecular system around its central element. We observed a positive correlation between $Z$ and $\tau_\mathsf{\,f}$ for random systems, including extreme cases and shuffled sequences. However, we have also found that the impact of the $Z$ parameter is comparable to that of the Brownian motion. In other words, although mean spreading times do increase with $Z$, we cannot reliably use $Z$ to predict the spreading time of a given molecular system. 

Finally, our results indicate that the distribution of the rescaled spreading times $\mathsf{t}=\tau_f/\langle \tau_\mathsf{\,f} \rangle$ is a universal function with a standard deviation that decreases like $\sim 1/\sqrt{N}$, independent of the other system parameters. An excellent fit is provided by a function suggested by the theory for single-particle first-passage problems.

\section{\label{sec:Acknowledgement}Acknowledgments}
{The MC simulations were performed on the computing resources provided by SHARCNET and Compute Canada.  GWS acknowledges the support of both the University of Ottawa and the Natural Sciences and Engineering Research Council of Canada (NSERC), funding reference number RGPIN/046434-2013 (Discovery Grant).}

\nocite{*}
\bibliography{aipsamp}

\begin{thebibliography}{19}%
\makeatletter
\providecommand \@ifxundefined [1]{%
 \@ifx{#1\undefined}
}%
\providecommand \@ifnum [1]{%
 \ifnum #1\expandafter \@firstoftwo
 \else \expandafter \@secondoftwo
 \fi
}%
\providecommand \@ifx [1]{%
 \ifx #1\expandafter \@firstoftwo
 \else \expandafter \@secondoftwo
 \fi
}%
\providecommand \natexlab [1]{#1}%
\providecommand \enquote  [1]{``#1''}%
\providecommand \bibnamefont  [1]{#1}%
\providecommand \bibfnamefont [1]{#1}%
\providecommand \citenamefont [1]{#1}%
\providecommand \href@noop [0]{\@secondoftwo}%
\providecommand \href [0]{\begingroup \@sanitize@url \@href}%
\providecommand \@href[1]{\@@startlink{#1}\@@href}%
\providecommand \@@href[1]{\endgroup#1\@@endlink}%
\providecommand \@sanitize@url [0]{\catcode `\\12\catcode `\$12\catcode `\&12\catcode `\#12\catcode `\^12\catcode `\_12\catcode `\%12\relax}%
\providecommand \@@startlink[1]{}%
\providecommand \@@endlink[0]{}%
\providecommand \url  [0]{\begingroup\@sanitize@url \@url }%
\providecommand \@url [1]{\endgroup\@href {#1}{\urlprefix }}%
\providecommand \urlprefix  [0]{URL }%
\providecommand \Eprint [0]{\href }%
\providecommand \doibase [0]{http://dx.doi.org/}%
\providecommand \selectlanguage [0]{\@gobble}%
\providecommand \bibinfo  [0]{\@secondoftwo}%
\providecommand \bibfield  [0]{\@secondoftwo}%
\providecommand \translation [1]{[#1]}%
\providecommand \BibitemOpen [0]{}%
\providecommand \bibitemStop [0]{}%
\providecommand \bibitemNoStop [0]{.\EOS\space}%
\providecommand \EOS [0]{\spacefactor3000\relax}%
\providecommand \BibitemShut  [1]{\csname bibitem#1\endcsname}%
\let\auto@bib@innerbib\@empty
\bibitem [{\citenamefont {Karolin}\ \emph {et~al.}(2022)\citenamefont {Karolin}, \citenamefont {Vilhelm}, \citenamefont {Sriram}, \citenamefont {Kevin},\ and\ \citenamefont {Fredrik}}]{ref:dorfman_review}%
  \BibitemOpen
  \bibfield  {author} {\bibinfo {author} {\bibfnamefont {F.}~\bibnamefont {Karolin}}, \bibinfo {author} {\bibfnamefont {M.}~\bibnamefont {Vilhelm}}, \bibinfo {author} {\bibfnamefont {K.}~\bibnamefont {Sriram}}, \bibinfo {author} {\bibfnamefont {D.~D.}\ \bibnamefont {Kevin}}, \ and\ \bibinfo {author} {\bibfnamefont {W.}~\bibnamefont {Fredrik}},\ }\href@noop {} {\bibfield  {journal} {\bibinfo  {journal} {Quarterly Reviews of Biophysics}\ }\textbf {\bibinfo {volume} {55(e12)}},\ \bibinfo {pages} {1} (\bibinfo {year} {2022})}\BibitemShut {NoStop}%
\bibitem [{\citenamefont {Schwartz}\ \emph {et~al.}(1993)\citenamefont {Schwartz}, \citenamefont {Li}, \citenamefont {Hernandez}, \citenamefont {Ramnarain}, \citenamefont {Huff},\ and\ \citenamefont {Wang}}]{ref:Schwartz1}%
  \BibitemOpen
  \bibfield  {author} {\bibinfo {author} {\bibfnamefont {D.~C.}\ \bibnamefont {Schwartz}}, \bibinfo {author} {\bibfnamefont {X.}~\bibnamefont {Li}}, \bibinfo {author} {\bibfnamefont {L.~I.}\ \bibnamefont {Hernandez}}, \bibinfo {author} {\bibfnamefont {S.~P.}\ \bibnamefont {Ramnarain}}, \bibinfo {author} {\bibfnamefont {E.~J.}\ \bibnamefont {Huff}}, \ and\ \bibinfo {author} {\bibfnamefont {Y.~K.}\ \bibnamefont {Wang}},\ }\href@noop {} {\bibfield  {journal} {\bibinfo  {journal} {Science}\ }\textbf {\bibinfo {volume} {262}},\ \bibinfo {pages} {110} (\bibinfo {year} {1993})}\BibitemShut {NoStop}%
\bibitem [{\citenamefont {Schwartz}\ \emph {et~al.}(1995)\citenamefont {Schwartz}, \citenamefont {Li}, \citenamefont {Hernandez}, \citenamefont {Ramnarain}, \citenamefont {Huff},\ and\ \citenamefont {Wang}}]{ref:Schwartz2}%
  \BibitemOpen
  \bibfield  {author} {\bibinfo {author} {\bibfnamefont {D.~C.}\ \bibnamefont {Schwartz}}, \bibinfo {author} {\bibfnamefont {X.}~\bibnamefont {Li}}, \bibinfo {author} {\bibfnamefont {L.~I.}\ \bibnamefont {Hernandez}}, \bibinfo {author} {\bibfnamefont {S.~P.}\ \bibnamefont {Ramnarain}}, \bibinfo {author} {\bibfnamefont {E.~J.}\ \bibnamefont {Huff}}, \ and\ \bibinfo {author} {\bibfnamefont {Y.~K.}\ \bibnamefont {Wang}},\ }\href@noop {} {\bibfield  {journal} {\bibinfo  {journal} {Nature genetics}\ }\textbf {\bibinfo {volume} {9}},\ \bibinfo {pages} {432} (\bibinfo {year} {1995})}\BibitemShut {NoStop}%
\bibitem [{\citenamefont {Reisner}\ \emph {et~al.}(2010)\citenamefont {Reisner}, \citenamefont {Larsen}, \citenamefont {Silahtaroglu}, \citenamefont {Kristensen}, \citenamefont {Tommerup}, \citenamefont {Tegenfeldt},\ and\ \citenamefont {Flyvbjerg}}]{ref:melt}%
  \BibitemOpen
  \bibfield  {author} {\bibinfo {author} {\bibfnamefont {W.}~\bibnamefont {Reisner}}, \bibinfo {author} {\bibfnamefont {N.~B.}\ \bibnamefont {Larsen}}, \bibinfo {author} {\bibfnamefont {A.}~\bibnamefont {Silahtaroglu}}, \bibinfo {author} {\bibfnamefont {A.}~\bibnamefont {Kristensen}}, \bibinfo {author} {\bibfnamefont {N.}~\bibnamefont {Tommerup}}, \bibinfo {author} {\bibfnamefont {J.~O.}\ \bibnamefont {Tegenfeldt}}, \ and\ \bibinfo {author} {\bibfnamefont {H.}~\bibnamefont {Flyvbjerg}},\ }\href@noop {} {\bibfield  {journal} {\bibinfo  {journal} {Proc Natl Acad Sci}\ }\textbf {\bibinfo {volume} {107}},\ \bibinfo {pages} {13294} (\bibinfo {year} {2010})}\BibitemShut {NoStop}%
\bibitem [{\citenamefont {Erik}\ \emph {et~al.}(2021)\citenamefont {Erik}, \citenamefont {Gaurav}, \citenamefont {Anna}, \citenamefont {Fredrik},\ and\ \citenamefont {Tobias}}]{ref:denseandsparse}%
  \BibitemOpen
  \bibfield  {author} {\bibinfo {author} {\bibfnamefont {T.}~\bibnamefont {Erik}}, \bibinfo {author} {\bibfnamefont {G.}~\bibnamefont {Gaurav}}, \bibinfo {author} {\bibfnamefont {J.}~\bibnamefont {Anna}}, \bibinfo {author} {\bibfnamefont {W.}~\bibnamefont {Fredrik}}, \ and\ \bibinfo {author} {\bibfnamefont {A.}~\bibnamefont {Tobias}},\ }\href@noop {} {\bibfield  {journal} {\bibinfo  {journal} {PloS one}\ }\textbf {\bibinfo {volume} {16}},\ \bibinfo {pages} {e0260489} (\bibinfo {year} {2021})}\BibitemShut {NoStop}%
\bibitem [{\citenamefont {Gruszka}\ \emph {et~al.}(2021)\citenamefont {Gruszka}, \citenamefont {Jeffet}, \citenamefont {Margalit}, \citenamefont {Michaeli},\ and\ \citenamefont {Ebenstein}}]{ref:Ebenst_review}%
  \BibitemOpen
  \bibfield  {author} {\bibinfo {author} {\bibfnamefont {D.}~\bibnamefont {Gruszka}}, \bibinfo {author} {\bibfnamefont {J.}~\bibnamefont {Jeffet}}, \bibinfo {author} {\bibfnamefont {S.}~\bibnamefont {Margalit}}, \bibinfo {author} {\bibfnamefont {Y.}~\bibnamefont {Michaeli}}, \ and\ \bibinfo {author} {\bibfnamefont {Y.}~\bibnamefont {Ebenstein}},\ }\href@noop {} {\bibfield  {journal} {\bibinfo  {journal} {Essays in Biochemistry}\ }\textbf {\bibinfo {volume} {65}},\ \bibinfo {pages} {51} (\bibinfo {year} {2021})}\BibitemShut {NoStop}%
\bibitem [{\citenamefont {Friedrich}\ \emph {et~al.}(2016)\citenamefont {Friedrich}, \citenamefont {Zec},\ and\ \citenamefont {Wang}}]{ref:Wang_review}%
  \BibitemOpen
  \bibfield  {author} {\bibinfo {author} {\bibfnamefont {S.~M.}\ \bibnamefont {Friedrich}}, \bibinfo {author} {\bibfnamefont {H.~C.}\ \bibnamefont {Zec}}, \ and\ \bibinfo {author} {\bibfnamefont {T.}~\bibnamefont {Wang}},\ }\href@noop {} {\bibfield  {journal} {\bibinfo  {journal} {LAB ON A CHIP}\ }\textbf {\bibinfo {volume} {16}},\ \bibinfo {pages} {790} (\bibinfo {year} {2016})}\BibitemShut {NoStop}%
\bibitem [{\citenamefont {Marzia}\ \emph {et~al.}(2025)\citenamefont {Marzia}, \citenamefont {Chandra}, \citenamefont {Ivy},\ and\ \citenamefont {Amit}}]{ref:Mapping_review}%
  \BibitemOpen
  \bibfield  {author} {\bibinfo {author} {\bibfnamefont {I.}~\bibnamefont {Marzia}}, \bibinfo {author} {\bibfnamefont {V.~N.}\ \bibnamefont {Chandra}}, \bibinfo {author} {\bibfnamefont {B.}~\bibnamefont {Ivy}}, \ and\ \bibinfo {author} {\bibfnamefont {M.}~\bibnamefont {Amit}},\ }\href {\doibase 10.1021/acs.analchem.4c06684} {\bibfield  {journal} {\bibinfo  {journal} {Analytical Chemistry}\ }\textbf {\bibinfo {volume} {97(16)}},\ \bibinfo {pages} {8641} (\bibinfo {year} {2025})}\BibitemShut {NoStop}%
\bibitem [{\citenamefont {Meng}\ \emph {et~al.}(2005)\citenamefont {Meng}, \citenamefont {Benson}, \citenamefont {Chada}, \citenamefont {Huff},\ and\ \citenamefont {Schwartz}}]{ref:channelCUtting}%
  \BibitemOpen
  \bibfield  {author} {\bibinfo {author} {\bibfnamefont {X.}~\bibnamefont {Meng}}, \bibinfo {author} {\bibfnamefont {K.}~\bibnamefont {Benson}}, \bibinfo {author} {\bibfnamefont {K.}~\bibnamefont {Chada}}, \bibinfo {author} {\bibfnamefont {E.~J.}\ \bibnamefont {Huff}}, \ and\ \bibinfo {author} {\bibfnamefont {D.~C.}\ \bibnamefont {Schwartz}},\ }\href@noop {} {\bibfield  {journal} {\bibinfo  {journal} {Proc Natl Acad Sci}\ }\textbf {\bibinfo {volume} {102}},\ \bibinfo {pages} {10012} (\bibinfo {year} {2005})}\BibitemShut {NoStop}%
\bibitem [{\citenamefont {Uppuluri}\ \emph {et~al.}(2021)\citenamefont {Uppuluri}, \citenamefont {Jadhav}, \citenamefont {Wang},\ and\ \citenamefont {Xiao}}]{ref:enzyme_cut1}%
  \BibitemOpen
  \bibfield  {author} {\bibinfo {author} {\bibfnamefont {L.}~\bibnamefont {Uppuluri}}, \bibinfo {author} {\bibfnamefont {T.}~\bibnamefont {Jadhav}}, \bibinfo {author} {\bibfnamefont {Y.}~\bibnamefont {Wang}}, \ and\ \bibinfo {author} {\bibfnamefont {M.}~\bibnamefont {Xiao}},\ }\href {\doibase 10.1021/acs.analchem.1c01373} {\bibfield  {journal} {\bibinfo  {journal} {Analytical Chemistry}\ }\textbf {\bibinfo {volume} {93(28)}},\ \bibinfo {pages} {9808} (\bibinfo {year} {2021})}\BibitemShut {NoStop}%
\bibitem [{\citenamefont {McCaffrey}\ \emph {et~al.}(2015)\citenamefont {McCaffrey}, \citenamefont {Sibert}, \citenamefont {Zhang}, \citenamefont {Zhang}, \citenamefont {Hu}, \citenamefont {Riethman},\ and\ \citenamefont {Xiao}}]{ref:enzyme_cut2}%
  \BibitemOpen
  \bibfield  {author} {\bibinfo {author} {\bibfnamefont {J.}~\bibnamefont {McCaffrey}}, \bibinfo {author} {\bibfnamefont {J.}~\bibnamefont {Sibert}}, \bibinfo {author} {\bibfnamefont {B.}~\bibnamefont {Zhang}}, \bibinfo {author} {\bibfnamefont {Y.}~\bibnamefont {Zhang}}, \bibinfo {author} {\bibfnamefont {W.}~\bibnamefont {Hu}}, \bibinfo {author} {\bibfnamefont {H.}~\bibnamefont {Riethman}}, \ and\ \bibinfo {author} {\bibfnamefont {M.}~\bibnamefont {Xiao}},\ }\href {\doibase 10.1093/nar/gkv878} {\bibfield  {journal} {\bibinfo  {journal} {Nucleic Acids Research}\ }\textbf {\bibinfo {volume} {44(2)}},\ \bibinfo {pages} {e11} (\bibinfo {year} {2015})}\BibitemShut {NoStop}%
\bibitem [{\citenamefont {Brochard}\ and\ \citenamefont {de~Gennes}(1977)}]{ref:deGennes_1}%
  \BibitemOpen
  \bibfield  {author} {\bibinfo {author} {\bibfnamefont {F.}~\bibnamefont {Brochard}}\ and\ \bibinfo {author} {\bibfnamefont {P.~G.}\ \bibnamefont {de~Gennes}},\ }\href {\doibase 10.1063/1.434540} {\bibfield  {journal} {\bibinfo  {journal} {J. Chem. Phys.}\ }\textbf {\bibinfo {volume} {67}},\ \bibinfo {pages} {52} (\bibinfo {year} {1977})}\BibitemShut {NoStop}%
\bibitem [{\citenamefont {Daoud}\ and\ \citenamefont {Gennes}(1977)}]{ref:deGennes_2}%
  \BibitemOpen
  \bibfield  {author} {\bibinfo {author} {\bibfnamefont {M.}~\bibnamefont {Daoud}}\ and\ \bibinfo {author} {\bibfnamefont {P.~G.~D.}\ \bibnamefont {Gennes}},\ }\href {\doibase 10.1051/jphys:0197700380108500} {\bibfield  {journal} {\bibinfo  {journal} {Journal de Physique}\ }\textbf {\bibinfo {volume} {38}},\ \bibinfo {pages} {85} (\bibinfo {year} {1977})}\BibitemShut {NoStop}%
\bibitem [{\citenamefont {Chubynsky}\ and\ \citenamefont {Slater}(2012)}]{ref:MykytaV2012}%
  \BibitemOpen
  \bibfield  {author} {\bibinfo {author} {\bibfnamefont {M.~V.}\ \bibnamefont {Chubynsky}}\ and\ \bibinfo {author} {\bibfnamefont {G.~W.}\ \bibnamefont {Slater}},\ }\href {\doibase 10.1103/PhysRevE.85.016709} {\bibfield  {journal} {\bibinfo  {journal} {Physical review. E}\ }\textbf {\bibinfo {volume} {85}},\ \bibinfo {pages} {016709} (\bibinfo {year} {2012})}\BibitemShut {NoStop}%
\bibitem [{\citenamefont {C.}(1993)}]{ref:FPTDist}%
  \BibitemOpen
  \bibfield  {author} {\bibinfo {author} {\bibfnamefont {B.~H.}\ \bibnamefont {C.}},\ }\href@noop {} {\emph {\bibinfo {title} {Random Walks in Biology.}}}\ (\bibinfo  {publisher} {Princeton University Press, Princeton, N.J.},\ \bibinfo {year} {1993})\BibitemShut {NoStop}%
\bibitem [{\citenamefont {Koki}\ \emph {et~al.}(2021)\citenamefont {Koki}, \citenamefont {Nobutoshi},\ and\ \citenamefont {Yo}}]{ref:review_nanochann}%
  \BibitemOpen
  \bibfield  {author} {\bibinfo {author} {\bibfnamefont {Y.}~\bibnamefont {Koki}}, \bibinfo {author} {\bibfnamefont {O.}~\bibnamefont {Nobutoshi}}, \ and\ \bibinfo {author} {\bibfnamefont {T.}~\bibnamefont {Yo}},\ }\href@noop {} {\bibfield  {journal} {\bibinfo  {journal} {Anal. Chem.}\ }\textbf {\bibinfo {volume} {93}},\ \bibinfo {pages} {332} (\bibinfo {year} {2021})}\BibitemShut {NoStop}%
\bibitem [{\citenamefont {Zhou}\ \emph {et~al.}(2009)\citenamefont {Zhou}, \citenamefont {Wei}, \citenamefont {Nguyen}, \citenamefont {Bechner}, \citenamefont {Potamousis}, \citenamefont {Goldstein}, \citenamefont {Pape}, \citenamefont {Mehan}, \citenamefont {Churas}, \citenamefont {Pasternak}, \citenamefont {Forrest}, \citenamefont {Wise}, \citenamefont {Ware}, \citenamefont {Wing}, \citenamefont {Waterman}, \citenamefont {Livny},\ and\ \citenamefont {Schwartz}}]{ref:repet_genome}%
  \BibitemOpen
  \bibfield  {author} {\bibinfo {author} {\bibfnamefont {S.}~\bibnamefont {Zhou}}, \bibinfo {author} {\bibfnamefont {F.}~\bibnamefont {Wei}}, \bibinfo {author} {\bibfnamefont {J.}~\bibnamefont {Nguyen}}, \bibinfo {author} {\bibfnamefont {M.}~\bibnamefont {Bechner}}, \bibinfo {author} {\bibfnamefont {K.}~\bibnamefont {Potamousis}}, \bibinfo {author} {\bibfnamefont {S.}~\bibnamefont {Goldstein}}, \bibinfo {author} {\bibfnamefont {L.}~\bibnamefont {Pape}}, \bibinfo {author} {\bibfnamefont {M.~R.}\ \bibnamefont {Mehan}}, \bibinfo {author} {\bibfnamefont {C.}~\bibnamefont {Churas}}, \bibinfo {author} {\bibfnamefont {S.}~\bibnamefont {Pasternak}}, \bibinfo {author} {\bibfnamefont {D.~K.}\ \bibnamefont {Forrest}}, \bibinfo {author} {\bibfnamefont {R.}~\bibnamefont {Wise}}, \bibinfo {author} {\bibfnamefont {D.}~\bibnamefont {Ware}}, \bibinfo {author} {\bibfnamefont {R.~A.}\ \bibnamefont {Wing}}, \bibinfo {author} {\bibfnamefont {M.~S.}\ \bibnamefont {Waterman}}, \bibinfo {author} {\bibfnamefont {M.}~\bibnamefont
  {Livny}}, \ and\ \bibinfo {author} {\bibfnamefont {D.~C.}\ \bibnamefont {Schwartz}},\ }\href@noop {} {\bibfield  {journal} {\bibinfo  {journal} {PLoS genetics}\ }\textbf {\bibinfo {volume} {5(11)}},\ \bibinfo {pages} {e1000711} (\bibinfo {year} {2009})}\BibitemShut {NoStop}%
\bibitem [{\citenamefont {Teague}\ \emph {et~al.}(2010)\citenamefont {Teague}, \citenamefont {Waterman}, \citenamefont {Goldstein}, \citenamefont {Potamousis}, \citenamefont {Zhou}, \citenamefont {Reslewic}, \citenamefont {Sarkar}, \citenamefont {Valouev}, \citenamefont {Churas}, \citenamefont {Kidd}, \citenamefont {Kohn}, \citenamefont {Runnheim}, \citenamefont {Lamers}, \citenamefont {Forrest}, \citenamefont {Newton}, \citenamefont {Eichler}, \citenamefont {Kent-First}, \citenamefont {Surti}, \citenamefont {Livny},\ and\ \citenamefont {Schwartz}}]{ref:4humangenome}%
  \BibitemOpen
  \bibfield  {author} {\bibinfo {author} {\bibfnamefont {B.}~\bibnamefont {Teague}}, \bibinfo {author} {\bibfnamefont {M.~S.}\ \bibnamefont {Waterman}}, \bibinfo {author} {\bibfnamefont {S.}~\bibnamefont {Goldstein}}, \bibinfo {author} {\bibfnamefont {K.}~\bibnamefont {Potamousis}}, \bibinfo {author} {\bibfnamefont {S.}~\bibnamefont {Zhou}}, \bibinfo {author} {\bibfnamefont {S.}~\bibnamefont {Reslewic}}, \bibinfo {author} {\bibfnamefont {D.}~\bibnamefont {Sarkar}}, \bibinfo {author} {\bibfnamefont {A.}~\bibnamefont {Valouev}}, \bibinfo {author} {\bibfnamefont {C.}~\bibnamefont {Churas}}, \bibinfo {author} {\bibfnamefont {J.~M.}\ \bibnamefont {Kidd}}, \bibinfo {author} {\bibfnamefont {S.}~\bibnamefont {Kohn}}, \bibinfo {author} {\bibfnamefont {R.}~\bibnamefont {Runnheim}}, \bibinfo {author} {\bibfnamefont {C.}~\bibnamefont {Lamers}}, \bibinfo {author} {\bibfnamefont {D.}~\bibnamefont {Forrest}}, \bibinfo {author} {\bibfnamefont {M.~A.}\ \bibnamefont {Newton}}, \bibinfo {author} {\bibfnamefont {E.~E.}\
  \bibnamefont {Eichler}}, \bibinfo {author} {\bibfnamefont {M.}~\bibnamefont {Kent-First}}, \bibinfo {author} {\bibfnamefont {U.}~\bibnamefont {Surti}}, \bibinfo {author} {\bibfnamefont {M.}~\bibnamefont {Livny}}, \ and\ \bibinfo {author} {\bibfnamefont {D.~C.}\ \bibnamefont {Schwartz}},\ }\href@noop {} {\bibfield  {journal} {\bibinfo  {journal} {Proc Natl Acad Sci}\ }\textbf {\bibinfo {volume} {107}},\ \bibinfo {pages} {10848} (\bibinfo {year} {2010})}\BibitemShut {NoStop}%
\bibitem [{\citenamefont {Ebert}\ \emph {et~al.}(2021)\citenamefont {Ebert}, \citenamefont {Audano}, \citenamefont {Zhu}, \citenamefont {Porubsky},\ and\ \citenamefont {Bonder}}]{ref:resolution_OM}%
  \BibitemOpen
  \bibfield  {author} {\bibinfo {author} {\bibfnamefont {P.}~\bibnamefont {Ebert}}, \bibinfo {author} {\bibfnamefont {P.}~\bibnamefont {Audano}}, \bibinfo {author} {\bibfnamefont {B.}~\bibnamefont {Zhu}, \bibfnamefont {Q.~Rodriguez-Martin}}, \bibinfo {author} {\bibfnamefont {D.}~\bibnamefont {Porubsky}}, \ and\ \bibinfo {author} {\bibfnamefont {M.~e.~a.}\ \bibnamefont {Bonder}},\ }\href@noop {} {\bibfield  {journal} {\bibinfo  {journal} {Science}\ }\textbf {\bibinfo {volume} {372}},\ \bibinfo {pages} {eabf7117} (\bibinfo {year} {2021})}\BibitemShut {NoStop}%
\end{thebibliography}%

\end{document}